\documentclass[11pt]{article}
\usepackage{latexsym}
\usepackage{graphicx}
\usepackage{amsthm}
\usepackage{amsfonts}
\usepackage{amsmath}
\usepackage{subcaption}
\usepackage{todonotes}
\usepackage{amssymb,amsxtra,mathtools,etoolbox}
\usepackage{graphicx,xparse,dirtytalk,color,soul,xcolor}
\usepackage{xspace}
\usepackage{comment}
\usepackage{todonotes}
\usepackage{authblk}

\usepackage[english, ruled, onelanguage, linesnumbered, vlined]{algorithm2e}

\usepackage{ulem}
\definecolor{turquoise}{rgb}{0.19, 0.84, 0.78}

\usepackage{tikz,pgf}
\usetikzlibrary{positioning,shapes.multipart, fit,backgrounds,calc,decorations.pathreplacing, graphs, graphs.standard}
\pgfdeclaredecoration{sl}{initial}{
  \state{initial}[width=\pgfdecoratedpathlength-1sp]{
     \pgfmoveto{\pgfpointorigin}
  }
  \state{final}{
     \pgflineto{\pgfpointorigin}
    }
}

\theoremstyle{plain}
\newtheorem{theorem}{Theorem}[section]
\newtheorem{lemma}{Lemma}[section]

\newtheorem{property}{Property}[section]

\theoremstyle{definition}
\newtheorem{definition}{Definition}[section]

\renewenvironment{proof}{\noindent{\bf Proof.} }{\null\hfill$\Box$\par\medskip}

\newenvironment{state}[2]{
\fbox{
\begin{minipage}{\columnwidth}
{\bf State} #1
\ \\
{\small #2}
\end{minipage}
}
}{\vspace{1em}}

\newcommand{\duv}[2]{(#1, #2)}

\newcommand{\Z}{\mathbb{Z}}

\newcommand{\dg}{\mathcal{D}}
\newcommand{\mg}{\mathcal{M}}
\newcommand{\tg}{\mathcal{G}}

\newcommand{\ch}{\mathcal{H}}
\newcommand{\gamegr}{\mathcal{C}}
\newcommand{\peri}[1]{\left(#1\right)^{*}}
\newcommand{\ag}{\mathcal{A}}
\newcommand{\agm}{\mathcal{A}^*}

\newcommand{\conf}{ }

\newcommand{\confgr}{\mathcal{CG}}
 \newcommand{\roo}{anchored}

\newcommand{\uni}{star}

\newcommand{\npg}[3]{\Gamma_{#1}\left(#2, #3\right)}
\newcommand{\npm}[3]{\Gamma_{#1}^{\mathrm{in}}\left(#2, #3\right)}

\newcommand{\hs}[1][]{%
    \ifthenelse{\isempty{#1}}%
        {h_u}%
        {h_{#1}}
}

\newcommand{\cs}{\sigma_c}
\newcommand{\rs}{\sigma_r}

\newcommand{\ccg}[1][]{%
  \ifthenelse{\isempty{#1}}%
  { augmented arena}%
  { augmented arenas}%
}

\newcommand{\cce}[1][]{%
  \ifthenelse{\isempty{#1}}%
  {shadow edge}%
  {shadow edges}%
  }
\newcommand{\ccc}[1][]{%
  \ifthenelse{\isempty{#1}}%
  { shadow corner}%
  { shadow corners}%
}
\newcommand{\ccv}[1][]{%
  \ifthenelse{\isempty{#1}}%
  {shadow cover}%
  {shadow cover}%
}
\newcommand{\pnei}[1][]{%
    \ifthenelse{\isempty{#1}}%
    {$p$-outneighbour}%
    {$p$-outneighbours}%
}

\newcommand{\eset}{\mathrm{\cal SE}}

\newcommand{\diff}{\mathrm{DIF}}

\newcommand{\algsize}[1]{\phi\left(#1\right)}

\newcommand{\bigo}[1]{\mathcal{O}\left(#1\right)}

\DeclarePairedDelimiterX{\set}[1]{\{}{\}}{\setargs{#1}}
\NewDocumentCommand{\setargs}{>{\SplitArgument{1}{;}}m}
{\setargsaux#1}
\NewDocumentCommand{\setargsaux}{mm}
{\IfNoValueTF{#2}{#1} {#1\,\delimsize|\,\mathopen{}#2}}%{#1\:;\:#2}

\DeclarePairedDelimiterX{\lis}[1]{[}{]}{\lisargs{#1}}
\NewDocumentCommand{\lisargs}{>{\SplitArgument{1}{;}}m}
{\lisargsaux#1}
\NewDocumentCommand{\lisargsaux}{mm}
{\IfNoValueTF{#2}{#1} {#1\,\delimsize|\,\mathopen{}#2}}%{#1\:;\:#2}

\newrobustcmd*{\myline}[1]{\tikz{\filldraw[draw=#1, line width = 3pt] (0,0) -- (0.5cm, 0);}}
\newrobustcmd*{\myfline}[1]{\tikz{\filldraw[draw=#1, line width = 0.2mm] (0,0) -- (0.5cm, 0);}}

\DeclarePairedDelimiter{\abs}{\lvert}{\rvert}
\makeatletter
\let\oldabs\abs
\def\abs{\@ifstar{\oldabs}{\oldabs*}}

\DeclarePairedDelimiter{\fl}{\lfloor}{\rfloor}
\makeatletter
\let\oldfl\fl
\def\fl{\@ifstar{\oldfl}{\oldfl*}}

\DeclarePairedDelimiter{\cl}{\lceil}{\rceil}
\makeatletter
\let\oldcl\cl
\def\cl{\@ifstar{\oldcl}{\oldcl*}}

\title{\bf Cops \& Robber on Periodic Temporal Graphs\thanks{In occasion of Ralf Klasing's 60th birthday, celebrating his  many outstanding contributions in the area of discrete algorithms, in particular  for  mobile agents and  dynamic networks, which are the topics of our paper.}}

\author[1]{Jean-Lou De Carufel}
\author[1]{Paola Flocchini}
\author[2]{Nicola Santoro}
\author[1]{Fr\'ed\'eric Simard}
\affil[1]{School of Electrical Engineering and Computer Science, University of Ottawa}
\affil[2]{School of Computer Science, Carleton University}

\date{}

\begin{document}
\maketitle

%===================
\begin{abstract}
We consider the {\em Cops and Robber} pursuit-evasion game when the edge-set of the graph is allowed to change in time, possibly at every round. Specifically, the game is played on an infinite periodic sequence $\tg = (G_0, \dots, G_{p-1})^*$ of graphs on the same set $V$ of $n$ vertices: in round $t$, the topology of $\tg$ is $G_i=(V,E_i)$ where $i\equiv t \pmod{p}$.

Concentrating on the case of a {\em single} cop, we provide a characterization of copwin periodic temporal graphs, establishing several basic properties on their nature, and extending to the temporal domain classical {\em C\&R} concepts such as {\em covers} and {\em corners}. Based on these results, we design an efficient algorithm for determining if a periodic temporal graph is copwin.

We also consider the case of $k>1$ cops. By shifting from a representation in terms of directed graphs to one in terms of directed multi-hypergraphs, we prove that all the fundamental properties established for $k=1$ continue to hold, providing a characterization of $k$-copwin periodic graphs, as well as a general strategy to determine if a periodic graph is $k$-copwin.

Our results do not rely on any assumption on properties such as connectivity, symmetry, reflexivity held by the individual graphs in the sequence. They are established for a {\em unified} version of the game that includes the standard games studied in the literature, both for undirected and directed graphs, and both when the players are fully active and when they are not. They hold also for a variety of settings not considered in the literature.
\end{abstract}

%===================
\newpage

%==============================
\section{Introduction}
%==============================

%---------------------------------------
\subsection{Framework and Background}
%---------------------------------------

%---------------------------------------
\subsubsection{Cops \& Robber Games}

  {\em Cops \& Robber} ({\em C\&R})  
 is a  pursuit-evasion game
 played in rounds on a finite  graph $G$ between 
 a set of $k\geq1$ cops and
 a single robber.
Before starting the game,  an initial position on the vertices of $G$ is chosen first by the cops, then 
by the robber. Then,  in each round, 
first the cops, then the robber,    move to  neighbouring vertices
 or (if allowed by the variant of the game) stay in the current location.
The game ends  if the cops {\em capture} the robber:
 at least one  cop moves to the  vertex currently occupied by the robber, 
in which case the cops have won. 
The robber wins by forever avoiding capture;
note that, in this case, the game never ends.

In the original  version, introduced 
 by  Quillot \cite{Qu78} and independently  by
Nowakowski and Winkler \cite{NoW83}, 
 the graph $G$ is connected and undirected,
there is a single cop  and,  in each round,  the
players  are allowed  not to move; it has then  been extended by 
Aigner and Fromme \cite{AiF84}  to permit multiple
cops.
This version, which we shall call  {\em standard},
is the most commonly investigated (see  \cite{BoN11}).

 Among the many variants of this game  (for a partial list, see \cite{BoM17,BoN11}),
two are of
particular interest to us. The first is the
 (much less investigated) natural generalization  when the graph $G$ 
is a strongly connected directed graph
\cite{DaGSS21,KhK3SV18,LoO17}; we shall refer to this version as {\em directed}.
Also of interest  is the variant,  called {\em fully active} (or {\em restless}), in which  the players 
must move in every round \cite{DaG22,GrKS20}; proposed for the standard  game, 
this variant  
can obviously be 
extended also to 
the directed version.

In the extensive   existing 
 research (see  \cite{BoN11} for a review), the main focus 
is on characterizing the class of {\em $k$-copwin} graphs;
i.e., those graphs 
where there exists a strategy  allowing  $k$ cops to capture
 the robber regardless of the latter's decisions.
 Related questions are to determine the minimum number  of cops 
capable of winning in $G$, called the {\em copnumber of} $G$,
or just to decide whether $k$ cops suffice.\
The  goal  underpinning this research is  the identification of
properties that allow the  characterization of 
graph classes  by means of the  copnumber of their members.

  The main  algorithmic question is the complexity of deciding whether
or not a graph is $k$-copwin as a function of the input parameters:
the number $n$ of vertices, the number $m$ of edges, and the
number $k$ of cops; particular attention has been given to the case 
when there is a single cop.
Currently, the most efficient  algorithm for 
 deciding whether
or not a graph is $k$-copwin in  the standard game
is $O(k n^{k+2})$  \cite{PePV22}, which yields
$O(n^3)$ for the case $k=1$.

In the  existing literature on the {\em C\&R} game, with only a couple of 
recent exceptions,
all results are based on a common assumption:   the
graph on which the game is played is {\em static};
that is, its link structure is the same in every round.

The question naturally arises:  what happens if
the {\em C\&R} game is played on a {\em time-varying} graph? More precisely, what happens if
the  link structure of the graph on which the
game is played  changes in time,
possibly in every round?
In addition to opening a new theoretical line of inquiry, 
this question is particularly relevant
in view of  the intense focus on  time-varying graphs	
in the last two decades
by researchers  from several fields.

%---------------------------------------
\subsubsection{Temporal Graphs}

The  extensive investigations on
properties and computatonal aspects of time-varying graphs 
 have been originally motivated 
by the development and increasing importance  of highly 
dynamic networks, where the topology 
is continuously changing. 
Such systems occur in a variety 
 of different settings,
 ranging from wireless ad-hoc networks to social networks.
 Various formal models have been advanced to describe  the dynamics
of these networks in terms of the dynamics of the changes in  
the graphs representing their topology (e.g., \cite{CaFQS12,HoS12,WeZF15}).

When  time is   {\em discrete}, as  in the {\em C\&R} games, 
the dynamics of these networks is usually
 described as an  infinite sequence
 $\tg = (G_0, G_1, \dots)$,  called
 {\em temporal graph} (or  {\em evolving graph}),  
 of  static  graphs $G_i =(V, E_i)$ on the same set $V$
of vertices; the graph $G_i$ is called {\em snapshot} (of $\tg$ at time $i$),
and the aggregate graph $G=(V, \cup_{i} E_i)$ is called the {\em footprint}
(or {\em underlying}) graph.
This model, originally suggested in \cite{HarG97} and independently  
proposed in \cite{Fe04}, has become the
 de-facto standard in the ensuing investigations.

All the studies are being carried out
under some assumptions restricting the arbitrariness
of the changes.
Some of these assumptions are on the ``connectivity"  of
the graphs $G_i$  in the sequence; they range from the (strong)
 {\em 1-interval connectivity} requiring every $G_i$ 
 to be connected (e.g., \cite{CaKNP15,CaKNP19,IlKW14,KuhLyO10}), to the 
 weaker  {\em temporal connectivity} allowing each  $G_i$ to be disconnected
 but requiring the sequence
 to be {\em connected over time}
(e.g., \cite{CaFMS14,GoSOKM21}). 

Another class  of assumptions is on the ``frequency"
of the existence of the links in the sequence. An important assumption in this class
 is {\em periodicity}: there exists a positive integer $p$ such that $G_i=G_{i+p}$
 for all $i\in\Z$  (e.g., \cite{FlMS13,IlW11,JatYG14}).
  Its importance 
follows from the fact that it models a condition occurring in a large  variety
of important settings, ranging from public transit networks
 to low-orbit satellite networks, to activity schedules.

There is a  large number of studies on  {\em mobile entities}
operating in temporal graphs, under different combinations of the above 
(and other) restrictive assumptions. Among them,
 computations include  
{\em graph exploration}, {\em dispersion}, and {\em gathering}
(e.g., \cite{AgMMS18,BouDP17,DiLDFS20,DiLFPPSV20,ErHK15,ErS18,GoSOKM21,GoSOKM18,IlKW14}; for a recent survey see  \cite{DiL19}).
Until very recently, 
none of these studies considered {\em C\&R} games.

%---------------------------------------
\subsection{C\&R in  Temporal Graphs}
%---------------------------------------

Conceptually, the extension of  a {\em C\&R} game
to a temporal graph $\tg = (G_0, G_1, \dots)$ is quite natural.
Initially, first the cops, then the robber, choose a starting position 
on the vertices  of $G_0$. At the beginning of round $t\geq 0$, 
the players are in $G_{t}$ and, 
after making their decisions and moves (according to the rules of the game),  
they find themselves in 
$G_{t+1}$ in the next round.
The game ends if and only if a cop moves to the  vertex 
currently occupied by the robber; in this case the cops have won.
The robber wins by forever preventing the cops from winning. 

%---------------------------------------
\subsubsection{Existing Results}

The extension of {\em C\&R} games to temporal graphs has been first investigated  by Erlebach and Spooner
 \cite{ErS20}. They considered the standard game with a single cop
 under the {\em periodic} frequency restriction, i.e., the sequence defining $\tg$ is periodic,
 each $G_i$ is undirected, players are allowed not to move,
 and $k=1$. For this setting, they presented an algorithm 
 that determines whether a periodic temporal graph is copwin
 in time $O(p\ n^3)$ where $p$ is the period of $\tg$ and $n$ the
 number of vertices. 
 They also stated that their algorithm can be
 extended to $k>1$ cops with a resulting $O(k\ p\ n^{k+2})$
 time complexity. In this pioneering study,  the results are
 obtained by  reformulating  the problem 
 in terms of a reachability problem and solving the latter; this, unfortunately
 does not provide insights on the special nature of the game when the graph changes in time.

Using the same  reduction to reachability games,  
Balev et al. \cite{BaLLPS20} studied the standard game 
in temporal graphs under the 
{\em 1-interval connectivity} restriction within a fixed 
time window,
and indicated how their algorithm can be
 extended to the case of $k>1$ cops.
 Also in this case, unfortunately, these results provide
 no insights on the  nature of the game in the temporal
 dimension.
 They also considered an ``on-line" version of the problem,
i.e., where the sequence of graphs is a priori unknown; these results
however are not relevant for the 
``full-disclosure" problem studied here.

Finally,
if the temporal graph is not given explicitly 
(i.e., as the sequence of snapshots), but only implicitly by
means of the Boolean 
 {\em edge-presence} function\footnote{The  edge-presence function $f(e,t)\in\{0,1\}$ indicates 
 for every $e$ in the footprint of $\tg$ and round $t$
 whether or not  edge $e$ is present in $G_t$   \cite{CaFQS12}.},
the problem
of deciding whether a single cop has a winning strategy
in the standard game on a periodic  temporal graph
has been shown to be  $NP$-hard  \cite{MoRW20}, answering a question raised in \cite{ErS20}.
It has been later shown that the problem  is $NP$-hard
even if the  footprint  $G=(V, \cup_{i} E_i)$ of the temporal graph
is a very simple graph (i.e., directed and undirected cycle) \cite{MoW21,EMSW24}.

With a different focus, 
the  study of the structural properties of copwin temporal graphs has just started, examining the relationship between the copnumber of
temporal graphs and that of its static components 
(e.g., snapshots and footprint) \cite{DeFSS23b,thesis}.

%--------------------------------------- 
 \subsection{Contributions}
 %---------------------------------------

 In this paper we focus on {\em C\&R} games in
 {\em periodic} temporal graphs,  concentrating 
 on the case of a {\em single} cop.

 We study the {\em unified} version of  the game 
defined as follow:
 In every round $i\geq 0$, the snapshot graph
 $G_i$ is directed and the players are
 restless.
 Let us point out that   the 
 standard  version, both in the original or
restless variant, as well as the non-restless directed version 
can  actually be redefined  as a restless game played on 
 (appropriately chosen) directed graphs:  
a pair of directed edges between a
pair of nodes corresponds to an unidirected link between them, and
 the presence of a self-loop at a node  
allows the players  currently there not to move to
a different node in the current round.
In other words, the {\em restless directed} version  of the game
includes all the  different versions  mentioned above. 
 In this {\em unified} version, for the {\em C\&R}  game to be
defined, and thus   {\em  playable},
the only requirement is that every node  in the graph must have 
an outgoing edge. 
In our investigation,  we will use this simple unified version.

 For the unified game, we provide a complete characterization
 of  copwin periodic temporal graphs, 
 establishing several basic properties on the 
 nature of a copwin  game in such graphs.
We do so by using a compact representation of 
periodic temporal graphs  as  static directed graphs,
we call arenas,  introducing the novel notion
of augmented arenas, and using these structures
to extend to the temporal domain classical concepts such as 
 {\em covers} and  {\em corners}.
 
These characterization results are {\em general}, 
in the sense that they  
do not rely on any  assumption on properties
such as connectivity, symmetry, reflexivity
held (or not held) by the individual snapshot graphs 
in the sequence.

Based on these results, we design an
algorithm for determining if a periodic temporal graph
is copwin, prove its correctness and analyze its time
complexity.   The total cost of the algorithm is $O(p\ n^2 + n m)$,
 where $m=\sum_{i\in\Z_p} |E_i|$ is the number of edges
 in the first $p$ snapshots. Thus,
 in periodic graphs with sparse snapshots. 
 the proposed algorithm terminates in $O(p\ n^2)$ time, which,
 in the static case, 
 becomes $O(n^2)$. 
Following the preliminary announcement of these results in \cite{DeFSS23}, it has been recently shown
that also the  reduction  to reachability games of \cite{ErS20}  can achieve  the same bound  \cite{EMSW24}.

We then consider the case  $k>1$ of multiple cops. 
By shifting from a representation in terms of directed graphs to 
one in terms of directed multi-hypergraphs,
it is possible to extend all the basic concept introduced for $k=1$.
Indeed, we prove that all the fundamental properties of augmented arenas
established for $k=1$ continue to hold in this extended setting,
providing a complete characterization of $k$-copwin periodic graphs.
These results lead directly to a solution strategy to determine if a periodic temporal graph is $k$-copwin; however, the immediate implementation of the strategy does not lead  to an improved time bound. 

All our results are established for the unified version of the game.
Therefore, all the characterization properties and algorithmic results
  hold  not only for  
 the standard  games studied in the literature but also for the much less studied directed games, 
  both when the players are restless and when they are not.
  They 
hold also for all those settings, not considered in the literature,
 where there is a  mix of nodes:  those where the players must leave and
those where  the players can wait; furthermore such a mix
might be time-varying (i.e., different in every round).

%==============================
\section{Terminology and Definitions}
\label{sec:definitions}
%==============================

%---------------------------------------
\subsection{Graphs and Time}
%---------------------------------------

%------------------
\subsubsection{Static Graphs}
\label{sec:static}

We denote  by $G = (V,E)$, 
or sometimes by $G = (V(G), E(G))$, 
the {\em directed graph}  with set of  vertices  $V$ 
and  set of edges $E \subseteq V\times V$.
A self-loop is an edge of the form
$\duv{u}{u}$; if $\duv{u}{u}\in E$ for all $u\in V$, then we will say that $G$ is 
{\em reflexive}.
If $\duv{v}{u}\in E$ whenever $\duv{u}{v}\in E$,
we will say that $G$ is {\em symmetric} (or {\em undirected}).

Given a vertex $u\in V(G)$, we  shall denote by $E^-(u)$  the set of edges incident on $u$, 
and by and by $E^+(u)$ those departing from $u$. 
A vertex $v\in V(G)$ is said to be a {\em source} if $E^-(u)=\emptyset$, and to be a {\em sink} 
if $E^+(u)=\emptyset$. $G$ is said to be {\em sourceless} if it contains no sources, and {\em sinkless} if
it contains no sinks.

Given a graph $G'$, if $V(G')\subseteq V(G)$ and $E(G')\subseteq E(G)$, then we say $G'$ is a {\em subgraph} of $G$ and write $G'\subseteq G$. A subgraph  $G'\subseteq G$  is \emph{proper},  written $G'\subset G$, if $G'\neq G$. 

For reasons apparent later,  we shall refer to a graph $G$ so defined as a  {\em static graph},  and say it is 
 {\em playable} if every vertex has at least one outgoing edge.

%------------------
\subsubsection{Temporal  Graphs}

A {\em time-varying graph} $\tg$ is a   graph  whose set of edges changes 
in time\footnote{The terminology in this section is mainly from \cite{CaFQS12}.}.
A  {\em temporal graph} is a time-varying graph where time is 
assumed to be discrete and to have a start, i.e., 
{\em time} is the set  $\Z^+$ of  positive integers including $0$.

A temporal graph $\tg$ is represented as an  infinite sequence  
$\tg = (G_0, G_1, \dots)$ of  static  graphs $G_i =(V,E_i)$ on the same set 
of vertices $V$; we shall denote by $n=|V|$ the number of  vertices.
The graph $G_i$ is called the \emph{snapshot} of $\tg$ 
\emph{at time} $i\in \Z^+$,  and the aggregrate graph $G = (V, \bigcup_{i} E_i)$
is 
called the {\em footprint} of $\tg$. 
A temporal graph  $\tg$ is said to be \emph{reflexive} if all its snapshots are
reflexive,  {\em symmetric} if all its snapshots are
symmetric, {\em sourceless} if all its snapshots are sourceless.

Given two verices $x, y\in V$, a  strict \emph{journey} (or  {\em temporal walk}),
from $x$ to $y$ starting at time $t$ 
is any finite sequence 
$\pi(x,y) = \langle (z_0, z_1),(z_1,z_2),\dots,(z_{k-1},z_k)\rangle$
where $z_0=x, z_k=y$, and   $(z_i, z_{i+1})\in E_{t+i}$ for $0 \leq i < k$.
In the following, for simplicity, we will omit the adjective ``strict".

A temporal graph $\tg$ is   \emph{temporally connected} if for any $u,v\in V$ and
any time $t\in \Z^+$  there is a  journey from $u$ to $v$ that starts at time $t$.
Observe that,  if $\tg$ is temporally connected, then  its footprint is strongly connected
even when all its snapshots are disconnected.
A temporal graph  $\tg$ is said to be \emph{always connected} (or
 $1$-{\em interval connected})  if all its snapshots are strongly connected.

A temporal graph  $\mathcal{G}$ is {\em periodic} if there exists a positive integer $p$ such that for all $i\in \Z^+$, $G_i = G_{i+p}$. If $p$ is the smallest such integer, then $p$ is called the {\em period} of $\tg$ and $\tg$ is said to be $p$-periodic.  We shall represent a $p$-periodic temporal graph $\tg$ as $\tg = \peri{G_0, \dots, G_{p-1}}$; all operations on the indices will be taken modulo $p$. 
An example of a temporal periodic graph $\mathcal{G}$ with $p=4$ is shown in
Figure~\ref{fig:example_tg};
observe that $\tg$  is temporally connected, however
most of its snapshots  are disconnected digraphs,
and none of them is strongly connected.

Let $\tg = \peri{G_0 \dots G_{p-1}}$ and $\ch = \peri{H_0 \dots H_{p-1}}$ be two temporal periodic graphs with the same period on the same set $V$ of vertices; 
we say $\ch$ is a \emph{periodic subgraph} of $\tg$, written $\ch\subseteq \tg$, if $H_i \subseteq G_i$ for every  $i\in \Z_p=\{0, 1,\ldots, p-1\}$. We shall denote by 
 $\ch\subset \tg$ the fact that $\ch$ is a {\em proper} subgraph of $\tg$, i.e.,
 $\ch\subseteq \tg$ but $\ch\neq \tg$.
Let us point out the  obvious but useful fact
that static graphs  are temporal periodic graphs with period $p=1$.

 In this paper we focus on {\em C\&R} games in
 {\em periodic} temporal graphs,  henceforth referred to simply as \emph{periodic graphs}, concentrating 
 on the case of a {\em single} cop, and then focusing on the
 case of {\em multiple} cops.

%------------------
\subsubsection{Arena}

Consider the following class of directed 
static graphs, we shall call
{\em arenas}.

\begin{definition}[Arena]\label{def:corners_arena}
Let $k\geq 1$ be an integer and  $W$ be a non-empty finite set.
An {\em arena}  of lenght  $k$ on $W$  is any
static directed graph $\mg = (\Z_k\times W, E(\mg))$ where
$E(\mg) \subseteq \{((i,w),([i+1]_{k}, w'))| i\in \Z_k\mbox{ and } w,w'\in W\}$,
and $[i]_k$ denotes $i$ modulo $k$.
\end{definition}  

\begin{figure}[h!]
\centering
\includegraphics[scale=1]{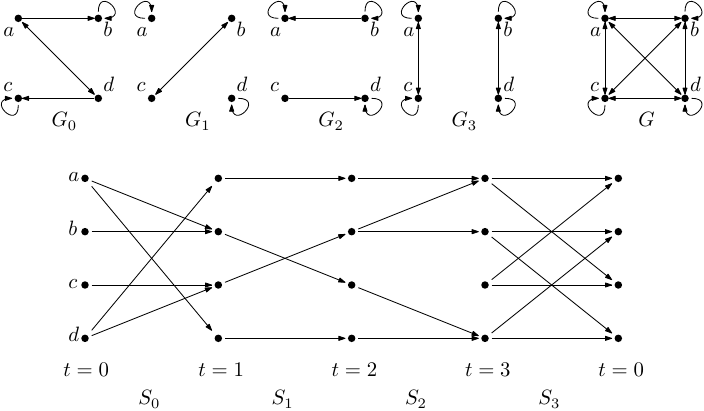}
\caption{A periodic graph  
 $\tg = \peri{G_0,G_1,G_2,G_{3}}$, its footprint $G$,
 and the corresponding arena.}
\label{fig:example_tg}
\end{figure}

A  periodic graph  
  $\tg = \peri{G_0, \dots, G_{p-1}}$
  with period $p$ and set of vertices $V$  
   has a unique correspondence with  the  arena
   $\dg  = (\Z_p\times V, E(\dg))$  where, for all  $i\in \Z_p$,  

 $$ ((i,u),([i+1]_p,v))\in E(\dg) \iff  \duv{u}{v}\in 
E_i,$$ 

\noindent  called the {\em arena of} 
 $\tg$. In particular,
the arena $\dg$ of $\tg$ 
explicitly preserves the snapshot structure of $\tg$:
  for all $i\in \Z_p$,
there is an obvious  one-to-one correspondence between 
the  snapshot $G_i$ of $\tg$ and the subgraph 
$S_i$ of $\dg$, 
called  \emph{slice}  (or {\em stage}), where 
$V(S_i) = \set{(i,v), v\in V}$
and  $E(S_i)= \{((i,u),([i+1]_p,v)) | (u,v)\in E_i)\}$.
An example of  a periodic graph $\tg$ and its  arena $\dg$ is shown in Figure~\ref{fig:example_tg}.
In the following, when no ambiguity arises,   $\dg$ shall indicate the
arena of $\tg$.

The vertices of an arena $\dg$ will be called {\em temporal nodes}.
Given a temporal node $(i,u)\in V(S_i)$ we shall denote by
  $N_i(u,\dg)$  the set of its outneighbours,
  and by
  $\Gamma_i(u,\dg)=\{v\in V | ([i+1]_p,v)\in N_i(u,\dg)\}$  
  the corresponding set of vertices in $G_{i}$.\color{black}
    
A temporal node $(i,u)\in V(S_i)$ is said to be a  \emph{\uni} if 
$\Gamma_i(u,\dg) =  V$. 
It is said to be {\em \roo}\ if there exists a journey from some
remporal node $(0,v)\in V(S_0)$ to $(i,u)$.

A  \emph{subarena}  of  $\dg  = (\Z_p\times V, E(\dg))$ is any
arena $\dg'  = (\Z_p\times V, E({\dg'}))$ where $E({\dg'}) \subseteq E({\dg})$; 
we shall denote by 
 $\dg' \subset \dg$ the fact that $\dg'$ is a  subarena of $\dg$ with
 $E({\dg'}) \subset E({\dg})$.

%---------------------------------------
\subsection{Cop \&  Robber Game   in Periodic Graphs}
%---------------------------------------

%----------------------- 
\subsubsection{Basics}

 The extension of the game from static to temporal graphs is quite natural.
Initially, first the cop, then the robber, chooses a starting position 
on the vertices  of $G_0$. Then, at each time $t\in \Z^+$,
first the cop, then the robber, moves to a vertex adjacent to its 
current position in $G_i$, where $i=[t]_p$.
Thus, in round $t$, the players are in $G_{[t]_p}$ and, 
after making their decisions and moves,  they find themselves in $G_{[t+1]_p}$ in the next round.
The game ends if and only if the cop moves to the  vertex 
currently occupied by the robber; in this case the cop has won. 
The robber wins by forever preventing the cop from winning. 

We consider the version of the game 
where all players are restless, i.e., they all move
 in each round. The only
requirement made by this version on $\tg$ is that it is {\em playable}: in each snapshot, every vertex  must have an outgoing edge; that is, every $G_i$ is sinkless. In the following we only consider
playable periodic graphs.
No other requirement 
such as connectivity, symmetry, and  reflexivity
is imposed on $\tg$.

We call this version of the game {\em unified}. Observe that the standard version,
both in the original or restless variant, as well as the non-restless directed version can actually
be redefined as a restless game played in this unified version: a pair of
directed edges between a pair of verices corresponds to an unidirected link between them, and
the presence of a self-loop at a vertex allows the players currently there not to move to a different
vertex in the current round. 

A play on the arena $\dg$ of  $\tg$ follows the play on $\tg$
in a direct obvious way: at each time $t\in\Z^+$, first the cop, then the robber, 
chooses  a new vertex in the out-neighbourhood 
of its  current position and moves there. 
The cop wins and the game ends if it  manages to move to 
a temporal node $([t+1]_p,u)$ while the robber is on $([t]_p,u)$. 
The robber wins by forever escaping capture from the cop, in which case the game never ends.

%----------------------- 
\subsubsection{Configurations and Strategies}

A \emph{configuration}  is 
a triple $\conf(t,c,r)\in \Z^+\times V\times V$,
 denoting the   position $c\in V$ of the cop and $r\in V$ of the robber
at the beginning of  round $t\in\Z^+$. 
Let $\confgr = (V(\confgr),E(\confgr))$ be 
 the infinite directed  graph, called {\em configuration graph} of  $\dg$, 
 describing all the possible configurations $\conf(t,u,v)$  
 and their temporal connection in $\dg$:
\begin{align*}
V({\confgr}) &=\{(t,u,v)  |\  t\in\Z^+;  ([t]_p,u),([t]_p,v)\in V(\dg)\} , \\
E({\confgr}) &=\{((t,u,v),(t+1,u',v')) |\  t\in\Z^+;  u\neq v;
 u'\in \npg{[t]_p}{u}{\dg}, v'\in \npg{[t]_p}{v}{\dg} \}.
\end{align*}
 Observe that
${\confgr}$ is acyclic;  the {\em source} nodes (i.e., 
the nodes with no in-edges) are those with $t=0$,
the {\em sink} nodes (i.e., the nodes with no out-edges)
are those with $u=v$.

A playing {\em strategy for the cop} is
any function $\cs: V({\confgr}) \rightarrow  V$
where, for every $(t,u,v)\in V({\confgr})$,
$\cs(t,u,v)\in \Gamma_{[t]_p}(u,\dg)$, and  $\cs(t,u,v)=u$ if $u=v$;
it  specifies where the cop should move  in round $t$ 
if the cop is at $([t]_p,u)$, the robber is at $([t]_p,v)$, 
and it is the cop's turn to move.
A playing strategy $\rs$
for the robber is defined in a similar way.

A configuration $\conf(t,u,v)$ is said to be {\em copwin}
if there exists a strategy $\cs$  such that, 
starting from  $\conf(t,u,v)$,
the cop  wins the game  regardless of the strategy $\rs$ of the  robber; such a strategy $\cs$ will be said to be 
 {\em copwin for}  $\conf(t,u,v)$.
 A strategy $\cs$ is said to be {\em copwin}  if 
there exists a temporal node $(0,u)$ such that
$\cs$ is winning for $\conf(0,u,v)$ for all $v\in V$.
 
If a copwin strategy exists, then $\tg$ and its arena $\dg$ are said to be {\em copwin}, else they are {\em robberwin}.

%==============================
\section{Copwin Periodic Graphs}
%==============================
\label{sect:CopwinPeriodicGraphs}

%---------------------------------------------------
\subsection{Preliminary}
\label{sec:corner}
%---------------------------------------------------

In the analysis of the standard game  played in a static graph, an important
role is played by the notions of   {\em corner}  node and its {\em cover}. 
The usual meaning  is that 
if the robber is on the corner, after the cop has moved to the cover, no matter where the robber plays, the robber gets captured by the cop {\em in the next round}. 

 In an arena $\dg$, the same meaning  is provided directly  by 
 the notions of ``temporal corner" and ``temporal cover".
 
 \begin{definition} (Temporal Corner and Temporal Cover)
 A temporal node $(t,u)$ in an arena $\dg$ is said to be a \emph{temporal corner} 
of  temporal node $(t+1,v)$ if 
 $u\neq v$ and
 \[ \npg{t}{u}{\dg}\subseteq \npg{t+1}{v}{\dg}. \] The  temporal node $(t+1,v)$
 is said to be a  \emph{temporal cover} of $(t,u)$.\\
\end{definition}

An important
 relationship between temporal corners and copwin arenas is the following.

\begin{lemma}\label{obs:temp_corner}
Every copwin arena contains a temporal corner.
\end{lemma}
\begin{proof}
Let $\dg$ be a copwin arena. Then there must exists a time $t$, 
a configuration $\conf(t,c,r)$,
and a move by the cop to a neighbouring temporal node
$(t+1,c')$ such that, regardless of where the robber 
moves in round $t$,  it is captured by the cop in its next move.  
In other words,  for every $w\in \npg{t}{r}{\dg}$, 
there exists a $z\in \npg{t+1}{c'}{\dg}$ such that $z = w$. 
This means that  $\npg{t}{r}{\dg}\subseteq \npg{t+1}{c'}{\dg}$;
that is,   $(t,r)$ is a temporal corner of $(t+1,c')$.
\end{proof}

This necessary condition, although important,  
provides only limited  indications on how  to solve the
characterization problem.

%---------------------------------------------------
\subsection{Augmented  Arenas and Characterization}
%---------------------------------------------------

The  crucial element in the characterization of copwin periodic graphs is
the notion of  {\em  augmented arena}.

\begin{definition}(Augmented Arena)\label{def:ccg}
Let $\dg$ be the  arena of $\tg$. 
An {\em  augmented arena}  $\ag$ of $\dg$ is 
an arena such that $\dg \subseteq \ag$ 
and, for each edge $((t,x),(t+1,y))\in E(\ag)$, 
the configuration $\conf(t,x,y)$
is winning for the cop in $\dg$.
\end{definition}

We shall refer to the edges of the augmented arena
$\ag$ of $\dg$ as {\em shadow edges}. Observe that, 
by definition, all edges  of $\dg$  are  shadow edges of $\ag$.

Let $\mathbb{A}(\dg)$
denote the set of augmented arenas  of $\dg$.
Observe that, by definition, $\dg\in \mathbb{A}(\dg)$.
Further observe the following:

\begin{property}
\label{prop-closed}
The partial order 
 $(\mathbb{A}(\dg),\subset)$ induced by edge-set inclusion 
on $\mathbb{A}(\dg)$ is a complete lattice. Hence  $(\mathbb{A}(\dg),\subset)$ has a maximum which we denote by $\agm$.
\end{property}
\begin{proof}
It follows from the fact that, by definition
of augmented arena, the set  $\mathbb{A}(\dg)$  
is closed under the union of augmented arenas.
\end{proof}

We have now the elements for the characterization of copwin periodic graphs.

\begin{theorem}\label{thm:univ_temp_node}
{\em (Characterization Property)}\\
An arena $\dg$ is copwin if and only  $\agm$ contains an \roo\ \uni.
\end{theorem}
\begin{proof}
({\bf  if})
Let $\agm$ contain an  \roo\ 
\uni\ $(t,u)$, $t\in \Z_p$.
By definition of \uni, $\npg{t}{u}{\agm} = V$. Thus, by definition of augmented arena, 
for every $v\in V$ the configuration $\conf(t,u,v)$ is copwin, i.e.,
there is a copwin strategy $\cs$ from $\conf(t,u,v)$. 

Since  $(t,u)$ is \roo, there exists a temporal node $(0,x)$ such that there is a journey $\pi$ from $(0,x)$ to  $(t,u)$.
Consider now the cop strategy   $\cs'$ of: (1)  initially positioning  itself on the 
 temporal node $(0,x)$, (2) then
 moving according to the journey $\pi((0,x),(t,u))$  and, once on  $(t,u)$, (3) 
following the copwin strategy $\cs$ from $\conf(t,u,w)$, where $w$ is the position of
the robber at the beginning of round $t$. This strategy  $\cs'$ is  
winning  for $\conf(0,x,v)$ for all $v\in V$; hence
 $\dg$ is copwin.

({\bf only if}) 
Let $\dg$ be copwin. We show that   there must exist an  augmented arena  $\ag$ of $\dg$ that contains an \roo\ \uni.
Since $\dg$ is copwin, by definition, there must exist some starting position
$(0,x)$ for the cop such that, for all positions initially chosen by the robber, the
cop eventually captures the robber. In other words, all the configurations
$\conf(0,x,v)$ with $v\in V$ are copwin; thus 
the arena $\ag$ obtained by adding to $E(\dg)$ the set of edges $\{((0,x),(1,v)) | v\in V\}$
is an augmented arena of $\dg$ and $(0,x)$ is an \roo\ 
\uni.
By  Property~\ref{prop-closed},  $E(\ag)\subseteq E(\agm)$ and the theorem follows.
\end{proof}

The characterization of copwin periodic graphs provided by
 Theorem~\ref{thm:univ_temp_node}
 indicates that,  to
determine whether or not an arena $\dg$ is copwin, 
 it suffices  to check whether $\agm$  contains a  \uni.

To be able to transform this fact into an effective solution procedure,  some additional concepts need to be introduced and
properties   established.

%---------------------------------------------------
\subsection{Shadow Corners and Augmentation}
%---------------------------------------------------

Other  crucial elements in the analysis of 
 copwin periodic graphs
are the concepts  of  corner and cover, introduced in Section~\ref{sec:corner}
for arenas,  now in the context  of  augmented
arenas.

\begin{definition}(Shadow Corner and Shadow Cover)
Let $\ag$ be an  augmented arena  of $\dg$. 
A temporal node $(t,u)$  is a \emph{shadow corner} of a temporal node $(t+1,v)$, 
with $v\neq u$,  if  
\[
\npg{t}{u}{\dg} \subseteq \npg{t+1}{v}{\ag}.
\]
\noindent The temporal node  $(t+1,v)$ will then be called the \emph{shadow cover} of $(t,u)$.

\end{definition}

\begin{figure}[h!]
\begin{center}
\includegraphics[scale=1]{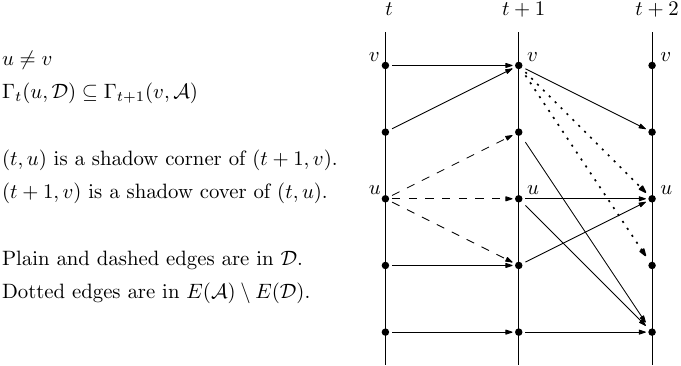}
\caption{Temporal node  $(t,u)$ is a shadow corner of $(t+1, v)$.}
\label{fig:covcor}
\end{center}
\end{figure}

\noindent By definition, any temporal corner is  a shadow corner,
and its temporal covers are shadow covers. An example is shown in Figure~\ref{fig:covcor};
the dashed links indicate the neighbours of node $(t,u)$ in $\dg$, while the dotted links indicate the edges to the neighbours of $(t+1, v)$ that exists in $\ag$ but not in $\dg$.

The role that shadow corners play with regards to
the set $\mathbb{A}(\dg)$ of  augmented arena  of  $\dg$
is expressed by the following.

\begin{theorem}\label{thm:ccg_add_edge} {\em (Augmentation Property)}\\
Let  $\ag\in \mathbb{A}(\dg)$,
$(t,x),(t,y)\in V(\dg)$
and   $z\in \npg{t}{x}{\dg}$.
If $(t,y)$ is a  shadow corner of $(t+1,z)$, then the arena $\ag' = \ag \cup \set{((t,x),(t+1,y))}$ is an  augmented arena  of $\dg$. 
\end{theorem}

\begin{proof}
Let $\ag$ be an  augmented arena  of $\dg$ and 
  let $(t,x),(t,y), (t+1,z)\in V(\dg)$ where  $z\in \npg{t}{x}{\dg}$ and
 $(t,y)$ is a  shadow corner of $(t+1,z)$.
The theorem  follows if  $((t,x),(t+1,y))$ is already an edge of $\ag$. 
Consider  the case where $((t,x),(t+1,y))\notin E(\ag)$. 
Since $(t,y)$ is a  shadow corner of 
$(t+1,z)$, then for every $w\in \npg{t}{y}{\dg}$  we have that
 $((t+1,z),(t+2,w))\in E(\ag)$, i.e.,  
$\conf(t+1,z,w)$ is winning for the cop.
 Since $z\in \npg{t}{x}{\dg}$, 
if the cop moves from $(t,x)$ to $(t+1,z)$ 
when the robber is on $(t,y)$, then regardless of the robber's move, 
the resulting configuration would be winning for the cop.
In other words, $\conf(t,x,y)$ is a winning configuration for the cop.
It follows that $\ag' = \ag \cup \set{((t,x),(t+1,y))}$ is an  augmented arena  of $\dg$.
\end{proof}

\noindent In other words,
 given an augmented arena, 
 by identifying a (still unconsidered) shadow corner and its covers,
new shadow edges may be 
determined and added to form a denser augmented arena.

%---------------------------------------------------
\subsection{Determining $\agm$}
%---------------------------------------------------

The properties expressed by Theorem~\ref{thm:ccg_add_edge},
in conjunction with that of Theorem~\ref{thm:univ_temp_node}, 
provide an algorithmic strategy to construct $\agm$:
start from an augmented arena;
determine new shadow edges; add them
to the set of shadow edges,  creating a denser 
augmented arena;
repeat this process  until  the current  augmented arena $\ag$ 
either contains an \roo\ \uni\
or  is $\agm$.

To be able to employ the above strategy, a condition is needed  to
determine if the current augmented arena  of $\dg$ is  indeed $\agm$.
This is provided by the following.

\begin{figure}[h!]
\begin{center}
\includegraphics[scale=1]{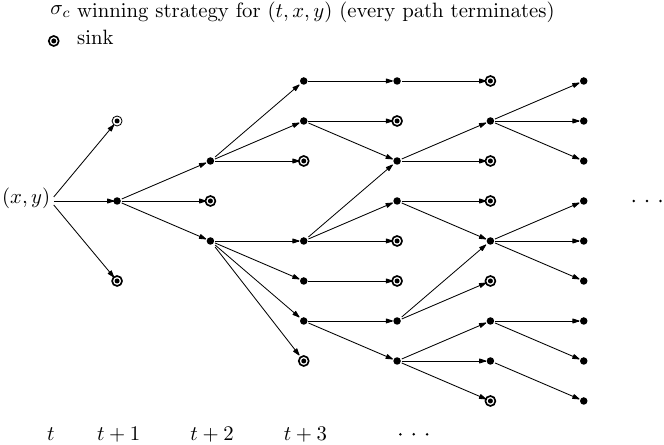}
\caption{The directed  acyclic graph $\gamegr$  of configurations 
induced by  $\cs$ starting from $\conf(t,x,y)$.}
\label{fig:maximal1}
\end{center}
\end{figure}

\begin{theorem}\label{thm:maximal}
{\em (Maximality Property)}\\
Let  $\ag\in \mathbb{A}(\dg)$. Then   $\ag=\agm$  if and only if, for every edge $((t,x),(t+1,y))\notin E(\ag)$, there exists no $z\in \npg{t}{x}{\dg}$ such that $(t,y)$ is a shadow corner of $(t+1,z)$.
\end{theorem}

\begin{proof} 
{\bf (only if)}\ By contradiction, let $\ag=\agm$ but there exists an edge $((t,x),(t+1,y))\notin E(\ag)$ and a temporal node $z\in \npg{t}{x}{\dg}$ such that $(t,y)$ is a shadow corner of $(t+1,z)$. By Theorem~\ref{thm:ccg_add_edge}, $\ag' = \ag \cup \set{((t,x),(t+1,y))}$ is an augmented arena of $\dg$; however, $E(\ag')$ contains one more edge than $E(\ag)$, contradicting the assumption that $\ag$ is maximum.
 
{\bf (if)}\ 
Let $\ag\neq\agm$; that is, there exists  $((t,x),(t+1,y))\in E(\agm)\setminus E(\ag)$. By definition, the configuration $\conf(t,x,y)$ is copwin; let $\cs$ be a copwin strategy for the configuration  $\conf(t,x,y)$, i.e., starting from  $\conf(t,x,y)$, the cop wins the game regardless of the strategy $\rs$ of the robber.
  
Let $\gamegr = (V(\gamegr),E(\gamegr))\subseteq \confgr$ be the directed  acyclic graph of configurations induced by $\cs$ starting from $\conf(t,x,y)$, and defined as follows:
(1) $\conf(t,x,y)\in V(\gamegr)$;
(2) if $\conf(t',u,v)\in V(\gamegr)$ with $t'\geq t$ and $u\neq v$, then, for all $w\in  \Gamma_{t'}(v,\dg)$, $\conf(t'+1,\cs(t'+1,u,v),w) \in V(\gamegr)$ and $(\conf(t',u,v), \conf(t'+1,\cs(t'+1,u,v),w)))\in E(\gamegr)$.
   
Observe that in $\gamegr$  there is only one source (or root) node, $\conf(t,x,y)$, and every  $\conf(t',w,w)\in V(\gamegr)$ is a sink (or terminal) node. Since $\cs$ is a winning strategy for the root, every node in $\gamegr$ is a copwin configuration, and every path from the root terminates in a sink node. See Figure~\ref{fig:maximal1}.
  
Partition $V(\gamegr)$ into two sets, $U$ and $W$ where $U=\{\conf(i,u,v) | ((i,u),(i+1,v))\in E(\ag)\}$ and $W= V(\gamegr) \setminus U$. Observe that  every sink  of $V(\gamegr)$ belongs to $U$; on the other hand, since $((t,x),(t+1,y))\notin E(\ag)$ by assumption, the root belongs to $W$ (see Figure~\ref{fig:maximal2}).
\begin{figure}[h!]
\begin{center}
\includegraphics[scale=1]{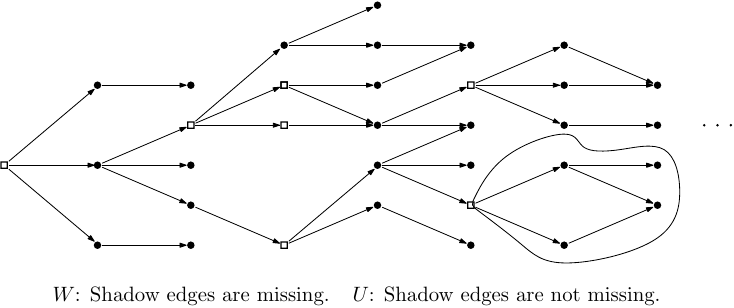}
\caption{The sets $W$ (${\scriptscriptstyle \square}$) and $U$ ($\bullet$).}
\label{fig:maximal2}
\end{center}
\end{figure}

Given a node $\kappa=\conf(i,u,v)\in V(\gamegr)$, let $\gamegr[\kappa]$ denote the subgraph of $\gamegr$ rooted in $\kappa$. \\

\noindent {\bf Claim.} {\em There exists $\kappa\in  V(\gamegr)$ such that
all nodes of $\gamegr[\kappa]$, except $\kappa$, belong to $U$. }\\
{\bf Proof of Claim.}
Let $P_0$ be the set of sinks of $\gamegr$. Starting from $k=0$, consider the set $P_{k+1}$ of all in-neighbours of any node of $P_{k}$; if $P_{k+1}$ does not contains an element of $W$, then increase $k$ and repeat the process. Since $(t,x,y)\in W$, this process terminates for some $k\geq 0$, and the Claim holds for every $\kappa\in P_{k+1} \cap W$. $\Box$\\

Let $(t',x',y')$ be a node of $V(\gamegr)$ satisfying the above Claim (see Figure~\ref{fig:maximal3a}). Thus $((t',x'),(t'+1,y'))\notin E(\ag)$ but, since $(t',x',y')$ is copwin, $((t',x'),(t'+1,y'))\in \agm$. By the Claim, all other nodes of $\gamegr[(t',x',y')]$ belong to $U$, in particular the set of nodes $\{(t'+1,w,z) | w = \cs(t',x',y'), z\in\Gamma_{t'}(y',\dg)\}$. This means that, for every $z\in\Gamma_{t'}(y',\dg)$, $(t'+1,w,z)\in E(\ag)$. In other words, $\npg{t'}{y'}{\dg}\subseteq \npg{t'+1}{w}{\ag}$; that is, $(t',y')$ is a shadow corner
of $(t'+1,w)$ (see Figure~\ref{fig:maximal3b}).
\begin{figure}[ht!]
\centering
\begin{subfigure}{0.375\textwidth}
    \centering
    \includegraphics[scale=1]{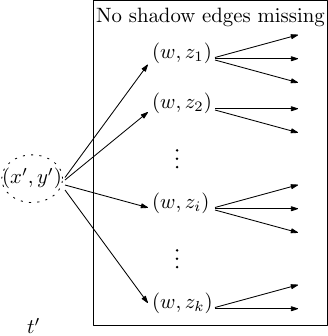}
    \caption{The situation in $\gamegr$. $(t',x',y')$ satisfies the Claim.}
    \label{fig:maximal3a}
\end{subfigure}
\hskip 0.1\textwidth
\begin{subfigure}{0.375\textwidth}
	\centering    
    \includegraphics[scale=1]{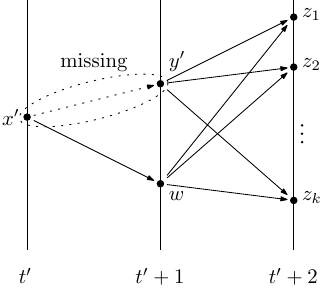}
    \caption{The situation in $\ag$. $(t',y')$ is a shadow corner
of $(t'+1,w)$.}
    \label{fig:maximal3b}
\end{subfigure}
\caption{The situation in $\gamegr$ and in $\ag$.}
\end{figure}

Summarizing: by assumption $\ag\neq\agm$; as shown, $((t',x'),(t'+1,y'))\in E(\agm)\setminus E(\ag)$, and $w\in \npg{t'}{x'}{\dg}$ is a shadow cover of $(t',y')$; that is, $\npg{t'}{y'}{\dg}\subseteq \npg{t'+1}{w}{\ag}$, concluding the proof of the {\bf if} part of the theorem.
\end{proof}

%==============================
\section{Algorithmic Determination}
%==============================

In this section we show that  the results established in the previous sections  
provide all the tools necessary to design an algorithm to  
determine whether or not a periodic graph $\tg$ is copwin. 
Furthermore,  
if $\tg$ is copwin, 
the algorithm can actually provide a winning cop strategy $\cs$.

%---------------------------------------
\subsection{Solution Algorithm}
%---------------------------------------

%---------------------------------------
\subsubsection{General Strategy}
\label{sec:general}

Given a periodic graph $\tg$, or equivalently
its arena $\dg$, to determine whether or not it is copwin, 
by Theorem~\ref{thm:univ_temp_node}, it is sufficient
to determine whether or not its maximal augmented arena $\agm$
contains an \roo\ \uni.  Hence, informally, a
basic solution approach  is to 
start from $\ag=\dg$, 
 repeatedly determine  a ``new" shadow edge 
(i.e., in $E(\agm)\setminus E(\ag)$)
using Theorem~\ref{thm:ccg_add_edge},
and consider the new augmented arena
obtained by adding such an edge.
This process is repeated  until  either the current  
augmented arena $\ag$ contains an \roo\ \uni, or no other ``missing'' shadow edge exists. In the former case, by Theorem~\ref{thm:univ_temp_node}, $\dg$ is copwin; in the latter case, by Theorem~\ref{thm:maximal}, the current augmented arena is $\agm$ and, if it does not contain an \roo\ \uni, $\dg$ is robberwin.

A general  strategy based on this approach 
 operates in a  sequence of iterations,
each composed of two operations: the examination
 of a   shadow edge, and the examination of  new shadow
 corners (if any) determined in the first operation. More precisely,
 in each iteration:
 (i)  A  ``new" (i.e., not yet examined) shadow edge   
 $e=((t,x),(t+1,y))$ 
 is examined 
to determine  if its presence 
transforms some temporal nodes  into new shadow corners of $(t,x)$.
(ii)  Each of these new shadow  corners is examined,  
determining   if its presence generates
new shadow edges. 
 By the end of the iteration, the  shadow edge $e$ 
 and the new shadow corners of $(t,x)$  examined 
 in this iteration are removed from consideration.
  This iterative process continues until there are
  no new shadow edges to be examined
  (i.e.  $\ag= \agm$) or  there is an \roo\ \uni\ in $\ag$.

 \begin{figure} 
\begin{center}
\fbox{
\begin{minipage}{5.5cm}
{\sc  General Strategy}
  \begin{tabbing} 
1. W\=hi\=le there  is a still unexamined shadow edge $e=((t,x),(t+1,y))$ in $\ag$ do:\\
2. \>  If there are still unexamined shadow corners   covered by $(t,x)$ then:\\
3. \> \>  F\=or each such shadow corner  $(t-1,z)$  do:\\
4. \> \>  \> If  th\=ere are new  shadow edges due to  $(t-1,z)$ then:\\
 5. \> \>  \>   \>            Add  them to  $\ag$ to be examined. \\
6. \> \>  \> Remove $(t-1,z)$ from consideration as a  shadow corners of $(t,x)$ (i.e., mark it as examined).\\
 7. \> Remove $e$ from consideration (i.e., mark it as examined).\\
 8. If there is an \roo\ \uni\ in $\ag$,  then $\dg$ is copwin else it is robberwin.
  \end{tabbing}
\end{minipage}
}
\end{center}
\caption{Outline of general strategy where the iterative process  terminates when $\ag= \agm$.} \label{fig:strategy}

\end{figure}

An outline of the strategy, where the
 iterative process is made to terminate when $\ag= \agm$,
  is shown in Figure~\ref{fig:strategy}.

%-------------------
\subsubsection{Algorithm Description}

Let us present the proposed algorithm, {\sc CopRobberPeriodic}, which follows
directly the general strategy described above to determine
whether or not an arena $\dg=((Z_p\times V),E(\dg))$ is copwin,
where  $V = \{v_1, \ldots, v_n\}$. 

We denote by $\ag$ the current augmented arena of $\dg$, by $A$ 
 its adjacency matrix, and by $A_t$  the adjacency matrix of slice 
 $S_t$ of  
$\ag$.
Auxiliary structures used by the algorithm include 
the queue $\eset$, 
of  the known shadow edges that have not been examined yet; 
a $n \times n$  Boolean matrix   $SE_t$ for each $t$,
 initialized to $A_t$,
used to indicate  shadow edges already known;
 a $n \times n$  Boolean matrix   $SC_t$ for each $t$,
 initialized to zero and
used to indicate the detected shadow corners;
more precisely, $B_t[x,y]=1$  indicates that $(t,x)$ 
has been determined to be a shadow corner of $(t+1,y)$,

The algorithm is composed of two phases:  {\em Initialization}, in which
all the necessary  structures are set up and
preliminary computations are performed;
and  {\em Iteration}, a repetitive process 
where  the two 
basic operations of the general strategy (described in Section~\ref{sec:general}) are performed
in each iteration: examination of a ``new" shadow edge
(to determine ``new" shadow corners generated by that edge) and 
examination of the ``new" shadow corners 
(to determine  ``new" shadow edges generated by that corner).

The  structure  used to  determine 
new shadow corners is  the set  
$ \{ \diff(t,x,y) : t\in\Z_p, x,y\in V\}$ of $n^2p$ Boolean  arrays  of dimension $n$.
For all $x,v\in V$ and $t\in\Z_p$, the value 
of the cell  $ \diff(t,x,y)[i]$ indicates whether
$$v_i \in \npg{t}{x}{\dg} \setminus \npg{t+1}{y}{\ag}$$ 
(in which case $\diff(t,x,y)[i] = 1$)
or $$v_i \in \npg{t}{x}{\dg} \cap \npg{t+1}{y}{\ag}$$ 
(in which case $\diff(t,x,y)[i] =  0$).  
Note that, if $v_i\notin \Gamma_t(x,\dg)$, the value of $ \diff(t,x,y)[i]$
is  left undefined; indeed,  the algorithm  only initializes and uses 
the $|\Gamma_t(x,\dg)|$ cells corresponding to the elements of $\Gamma_t(x,\dg)$;
 we shall call those cells the {\em core} of $ \diff(t,x,y)$.
 
The algorithm   also maintains a variable 
 $\algsize{\diff(t,x,y)}$ indicating the current number of core cells with value ``1" in array $\diff(t,x,y)$; this variable is initialized to $|\Gamma_t(x,\dg)|$.
Observe that, by definition of $\diff(t,x,y)$, 
 $\algsize{\diff(t,x,y)}=0$ iff $(t,x)$ is a shadow corner of $(t+1,y)$.

In   each iteration of the {\em Iteration} phase, a
new shadow edge is  taken from $\eset$, added to the augmented
arena $\ag$, and examined. The examination of a shadow edge $((t,x), (t+1,y))$ involves 
(i) the update of $\diff(t-1,z,x)[y]$ for any   
 in-neighbour  $(t-1,z)$,  in $\dg$,  of $(t,y)$  and, for any such in-neighbour, 
 (ii) the test to see if the presence of the edge $((t,x), (t+1,y))$ in the augmented arena has created new shadow corners among such in-neighbours\footnote{Such
 would be any   $(t-1, z)$   for which the update has resulted in an array $\diff(t-1,z,x)$ that contains only zero entries.}.
If new shadow corners exist, they may in turn have created new shadow edges originating from the in-neighbours, in $\dg$, of
$(t,x)$. In fact, any in-neighbour $(t-1,w)$ of $(t,x)$ such that   $((t-1,w),(t,z))$   is not already in the augmented arena is a new shadow edge:  
a move of the cop from $(t-1,w)$ to  $(t,x)$ is fatal for the robber wherever it goes;
in such a case, the algorithm then  adds $((t-1,w),(t,z))$ to $\eset$.

The pseudo code of the algorithm is shown in Algorithm~\ref{algo}.
 Observe that, in the algorithm,   
 the set of in-neighbours of a temporal node $(t,v)$ is denoted by $\npm{t}{v}{\dg}=\{z\in V | ((t-1,z),(t,v))\in E(\dg)\}$, 

 Not shown are   several very low level (rather trivial) 
 implementation details. 
 These include, for example, the fact that
  the core cells of $\diff(t,x,y)$ are connected through a 
  doubly linked list, and that, for efficiency reasons, we  
  also maintain two additional doubly linked lists: one  
  going through the  core cells of the array containing ``1", 
  the other linking the core cells containing ``0".

\begin{algorithm}
\DontPrintSemicolon
\KwIn{Arena $\dg=(\Z_p\times V,E(\dg))$, with $V = \{v_1, \ldots, v_n\}$}
{\em Initialization}\\
 $\ag := \dg$\\
$SE:=A$ \\ 
$\eset = \emptyset$\\
$SC:=${\em Zero} \qquad /*\quad a table of $p$ zero matrices, each of size $n\times n$\quad*/\\
 
\ForEach{$t\in\Z_p$, $u,v\in V$}{
      $\algsize{\diff(t,u,v)}:=  |\Gamma_t(u,\dg)|$\\        
       \ForEach{$w \in \npg{t}{u}{\dg}$}{
            \If{$A_{t+1}[v,w] = 1$}{
                    $\diff(t,u,v) [w] := 0$\\
                    $\algsize{\diff(t,u,v)}:=  \algsize{\diff(t,u,v)}-1$\\   
                    \If{$\algsize{\diff(t,u,v)} = 0$ \emph{\textbf{and}} $SC_{t}[u,v] = 0$} {                               
                         $SC_{t}[u,v] := 1$\\

                          \ForEach{$z\in \npm{t+1}{v}{\dg}$} {
                                   \If{$SE_{t}[z,u] = 0$}  {  
                                        $SE_{t}[z,u] := 1$\\
                                        $\eset \leftarrow ((t,z),(t+1,u))$\\
                                     }
                           }
                    }
             }
            \Else{ $\diff(t,u,v)[w] := 1$ }
        }
    
} 
{\em Iteration}\\
\While{$\eset\neq \emptyset$}{\label{line:while_loop_start}
    $((t,x),(t+1,y))  \leftarrow \eset$\\
    $A_t(x,y) := 1$ \\
 
        \ForEach{$z\in \npm{t}{y}{\dg}$}{

  \If{   $\diff(t-1,z,x)[y] = 1$ } 
      { 
      $\diff(t-1,z,x)[y] := 0$  \\        
      $\algsize{\diff(t-1,z,x)} := \algsize{\diff(t-1,z,x)}-1$ 
            }
            \If{$\algsize{\diff(t-1,z,x)} = 0$ \emph{\textbf{and}} $SC_{t-1}[z,x] = 0$}{\label{line:if_conf_visited}
              $SC_{t-1}[z,x] := 1$\\ 
                \ForEach{$w\in \npm{t}{x}{\dg}$}{
            \If{$SE_{t-1}[w,z] = 0$}{ \label{line:if_conf_visited_cop}
                $SE_{t-1}[w,z] := 1$\\
                $\eset \leftarrow ((t-1,w),(t,z))$\\

            }
           
        }
                }
            }

} 
\lIf{$\ag$  {\em contains an \roo\ \uni}} {$\dg$ is copwin} 
 \lElse {$\dg$  is robberwin.} 
\caption{{\sc CopRobberPeriodic}}
\label{algo}
\end{algorithm}

\newpage

%-------------------
\subsection{Analysis}
\label{analysis}
%-------------------

%-------------------
\subsubsection{Correctness}
%-------------------

Let us prove the correctness of Algorithm {\sc CopRobberPeriodic}.
Let $\dg=(\Z_p\times V,E(\dg))$ be the
 arena of a $p$-periodic graph with $n=|V|$ and $m= |E(\dg)|$.

\begin{lemma}\label{lem:terminate}
Algorithm {\sc CopRobberPeriodic} terminates after at most $|E(\agm)| - |E(\dg)|$
iterations.
 \end{lemma}
\begin{proof}
In the {\em Initialization} phase, all of the $m=|E(\dg)|$ 
 edges of $\dg$ are
examined, and their entry in the shadow edge matrix $SE$  is set to 1
(Line 3). Any new shadow edge discovered in this phase
is inserted in $\eset$ (Line 17).

Observe that, when a shadow edge $e$ is inserted in $\eset$, 
its entry in the shadow edge matrix $SE$  is set to 1
(this is done in Line 16 for the  edges of $\dg$, 
and in Line 32 for the others); this means that, once extracted and examined, 
$e$  will fail the test of Line 15 (or the test of Line 31) in any subsequent
iteration  and, therefore, it will never be inserted in $\eset$ again.  
Since only one shadow edge is
extracted from $\eset$ and examined in each iteration,
the number of iterations is at most the
total number  $|E(\agm)| - |E(\dg)|$ 
of shadow edges not originally in $\dg$.
 \end{proof}

Given an augmented arena $\ag$
and  a shadow   edge $e=((t,x),(t+1,y))\in E(\agm)\setminus E(\ag)$,
we shall say that $e$ is an {\em implicit} shadow
edge of $\ag$ if  there exists $z\in \npg{t}{x}{\dg}$
such that $(t,y)$ is a  shadow corner of $(t+1,z)$ in $\ag$. 

\begin{lemma}\label{lem:init}
At the end  of the {\em Initialization} phase:\
$(i)$ for all and only the temporal corners  $(t,x)$ of $(t+1,y)$ in $\dg$,
$SC_t[x,y]=1$ and $\algsize{\diff(t,x,y)} = 0$;
$(ii)$  all  implicit shadow edges of $\dg$  are in
$\eset$; furthermore, the entry in  $SE$ of all  edges of $\dg$ and 
 implicit shadow edges of $\dg$, is $1$.
\end{lemma}

\begin{proof}

({\bf i}) Observe that, in the  {\em Initialization} phase,
by construction,
 $\forall t \in\Z_p$, $\forall (t,u), (t+1,v) \in V(\dg)$,
and $\forall w\in \Gamma_t(u,\dg)$,
$\diff(t,u,v)$
 is initialized so that  $\diff(t,u,v)[w] = 1$
if and only if $w \in \npg{t}{u}{\dg} \setminus \npg{t+1}{v}{\dg}$.
Every time it is determined that $\diff(t,u,v)[w] = 0$,
the counter $\algsize{\diff(t,u,v)}$, initialized to
$|\Gamma_t(u,\dg)|$, is decreased by one; thus, by definition,
$\algsize{\diff(t,u,v)} = 0$ if and only if $(t,u)$ is a
temporal corner of $(t+1,v)$; in such a case $SC_t[u,v]=1$.
Recall that, by definition, all temporal corners are also
shadow corners. 

({\bf ii}) First observe that all edges of $\dg$ are by definition shadow edges,
and that their corresponding entry in the shadow edges matrix $SE$ is set to 1
(Line  3); hence, for them, the lemma holds.

Let us now consider the implicit shadow edges of $\dg$. Recall that, by Theorem~\ref{thm:ccg_add_edge}, given a  shadow corner 
$(t,u)$  of $(t+1,v)$, any edge  originating from an  in-neighbour $(t,z)$ of $(t+1,v)$  and terminating in  $(t+1,u)$ is a shadow edge (Lines 14-17); hence,
it is immediate to identify the implicit shadow edges corresponding
to a given shadow corner.  
An implicit shadow edge (i.e., one whose entry in $SE$ is $0$), once  identified,
 is   added to the queue $\eset$, and the corresponding entry in the shadow edges matrix is set to 1.
Since,  by part (i) of this Lemma, all  the shadow corners present in $\dg$ are identified, it follows that  all the implicit  shadow edges
  are queued in $\eset$ and their entry in $SE$ is set to 1.
\end{proof}

Let us consider the {\em Initialization} phase as iteration $0$
of the {\em Iteration} phase; hence, the entire algorithm can be
viewed as a sequence of iterations.
Denote by $\ag_j$  the augmented arena at the beginning
of the $j$-th iteration, with  $\ag_0 = \dg$.
We now show that, at the beginning of iteration $j$, 
all shadow corners of $\ag_{j-1}$
have been examined and  all implicit shadow edges of $\ag_{j-1}$
 are in $\eset$.

\begin{lemma}\label{lem:iteration}
At the beginning of iteration $j>0$:
\begin{description}
\item {\em (a)} $\algsize{\diff(t,x,y)} = 0$  if and only if
 $(t,x)$ is a shadow corner of $(t+1,y)$  in $\ag_{j-1}$;
 furthermore, in such a case,
$SC_t[x,y]=1$.
\item   {\em (b)} 
$\eset$ contains all the  
implicit shadow edges of $\ag_{j-1}$;
furthermore, in  $SE$, the entry  of the  edges of $\ag_{j-1}$ and 
of the implicit shadow edges of $\ag_{j-1}$ is $1$.
\end{description}
\end{lemma}

\begin{proof}
By induction on $j$.  Observe that, when $j=1$, both statements of the lemma follow directly from Lemma~\ref{lem:init}.
 Let they hold for $j\geq 1$; we now prove that they hold for $j+1$.
 
 Let $e_j = ((t,x),(t+1,y))$ be the shadow edge extracted from $\eset$ and examined in
iteration $j$. This edge is added to $\ag_{j-1}$ (Line 23), which is thus
transformed into $\ag_j$. 
This addition, which modifies only  the out-neighbourhood of $(t,x)$,
might create new shadow corners of $(t,x)$ among the  in-neighbours of $(t,y)$.
The algorithm therefore checks the set $\Gamma^{in}_t(y,\dg)$ to verify if
this has happened (Lines 24-33). This is done by considering, for each element $(t-1,z)$
of that set, the  entry $\diff(t-1,z,x)[y]$.

If  $\diff(t-1,z,x)[y]=1$, then the 
 shadow edge $e_j$ was one of those missing edges;
hence,  in that case (Lines 25-27)
the value of $\diff(t-1,z,x)[y]$ is set to zero 
and $\algsize{\diff(t-1,z,x)}$ is decreased by one.
If now $\algsize{\diff(t-1,z,x)}$ becomes $0$, then 
 $(t-1,z)$ is a shadow corner  of $(t,x)$  in $\ag_j$
 and $SC_{t-1}[z,x]$ is set to $1$ (Line 29).
 This means that, at the end of this iteration,  all shadow corners of 
 $\ag_j\setminus \ag_{j-1}$ have their entry in  $SC$ set to $1$
  and the corresponding entry in $\algsize{\cdot}$  set to $0$; thus, by
the  inductive hypothesis on the shadow corners of $\ag_{j-1}$, 
 statement  {\em (a)}  of the lemma holds for iteration $j$.

Any new shadow corner $(t-1,z)$  of $(t,x)$, 
created by the addition of $e_j$,
 might  in turn have created  new implicit
shadow edges in $\ag_j$. 
By Theorem~\ref{thm:ccg_add_edge}, any 
 edge  originating from 
 an  in-neighbour $(t-1,w)$ of $(t,x)$  and terminating in  
 the shadow corner $(t,z)$ 
 is a shadow edge (Lines 30-33). Let $P$ denote the set
 of these  shadow edges;  among them, the only implicit ones
 for $\ag_j$ are,  by inductive hypothesis, the ones whose entry in $SE$
 was $0$ in $\ag_{j-1}$. Any such implicit shadow edge
 is thus identified, added to the queue $\eset$, and the corresponding entry 
 in $SE$ is set to 1.
Since,  by part  ({\em a}) of this Lemma, all  the shadow corners present in $\ag_j$ are identified, it follows that  all the new implicit  shadow edges of $\ag_j$
  are queued in $\eset$ and their entry in $SE$ is set to 1.
  Thus, by
 inductive hypothesis on the shadow edges of $\ag_{j-1}$, 
 statement  ({\em b})   of the lemma holds for iteration $j$.
\end{proof}

\begin{theorem}\label{prop:alg_cubic_correct}
Algorithm {\sc CopRobberPeriodic} correctly determines 
whether or  not an arena $\dg$ is copwin.
\end{theorem}
\begin{proof}
By Lemma~\ref{lem:terminate}, the algorith terminates after
a finite number $q\geq 1$  of iterations, when
$\eset$ becomes empty and no other shadow edges are added
to it during the iteration. By Lemma~\ref{lem:iteration}, the fact that  $\eset=\emptyset$ 
at the end of the iteration means that in $\ag_q$ there are 
no implicit shadow corners identified in previous iterations; furthermore,
 during this iteration, regardless of
the new shadow corners found and examined, no implicit shadow corners were found.
In other words, the set $E(\agm) - E(\ag_q)=\emptyset$, i.e., $\ag_q=\agm$.
Hence, by Theorem~\ref{thm:maximal}, the  test in the last operation of the algorithm 
(Lines 34-35)
determines correctly whether or not $\dg$ is copwin.
 \end{proof}

%-------------------
\subsubsection{Complexity}
%-------------------

Let us analyze the cost of Algorithm {\sc CopRobberPeriodic}.
Given $\dg=(\Z_p\times V,E(\dg))$, let $m_i$ denote the number of edges of slice $S_i$ of $\dg$, $i\in Z_p$, and $m= |E(\dg)| = \sum_{i=0}^{p-1} m_i$  the total number of edges of $\dg$. As usual, $n = |V|$.

\begin{theorem}\label{thm:complexity}
Algorithm {\sc CopRobberPeriodic}  determines in  time $O(n^2 p + n m)$
whether or not  $\dg$ is copwin. 
\end{theorem}

\begin{proof} We first derive the cost of the
{\em Initialization} phase.  Observe that the initialization of $\ag$, 
$SE$, $SC$ (Lines 2-4)  can be performed with $O(n^2 p)$ operations.
Line 7 will be executed $n^2 p$ times.
The cost of the initialization of $\diff$  and of $\algsize{\diff}$ (Lines 6-13,18-19),
which  includes the update of some entries of $SC$, plus the cost of the 
initialization of $\eset$ (Lines 14-17),  which includes the update of some entries of $SE$,
 require   at most
 
 $$O(n^2p)+\sum_{i\in \Z_p,u\in V}
 O(|\Gamma_{i}(u, \dg)|) + \sum_{i\in \Z_p,v\in V} O(|\Gamma^{in}_i(v,\dg)|) =
O(n^2p) + \sum_{i=0}^{p-1} O(n (m_i + m_{i-1})) $$

\noindent operations, which sums up to $O(n^2 p + n m)$ operations for the {\em Initialization} phase.

Let us consider now the {\em Iteration} phase. The while loop will be repeated
until in the current augmented arena $\ag$ there are no more
shadow edges to be examined (i.e.  $\ag=\agm$).
By Lemma~\ref{lem:terminate},
the total number of iterations is  $|E(\agm)| -  |E(\dg)| \leq n^2 p - m$.
 Further observe that every operation performed during an iteration requires constant time. 

In each iteration, two processes are being carried out. 

The first process (Lines 24-27) is the determination of all
new shadow corners (if any) of $(t,x)$ created  by (the addition of) the
shadow edge $((t,x),(t+1,y))$ being examined.
The total cost of this process in this iteration is at most two operations for each
in-neighbour of $(t,y)$, 
i.e.,  at most $2 c_1 |\Gamma^{in}_t(y,\dg)|$, where $c_1\in O(1)$ is
the constant cost of  performing a single operation in this process. 

This process is repeated in all iterations, each time with a different
shadow edge being examined. 
Thus,
the cost of $2 c_1 |\Gamma^{in}_t(y,\dg)|$ will be incurred
for all $((t,x),(t+1,y))\in E(\agm)$; that is, at most $n$ times.
Summarizing, for each $y\in V, t\in \Z_p$ this process costs
$2 c_1 n |\Gamma^{in}_t(y,\dg)|$.
Hence the total cost of this process over all iterations is
$$
\sum_{y\in V, t\in \Z_p}
2\ c_1\ n\  |\Gamma^{in}_t(y,\dg)|
= 
4\ c_1\ n\ \sum_{t=0}^{p-1}
 m_t 
=
O(n m).$$

The second process, to be performed only if new shadow corners of
$(t,x)$ have been found in the first process,
 is the determination (Lines 28-33) of all the
new shadow edges (if any)  created  by the found
new shadow corners, and their addition to $\eset$.
The cost of this process for a new shadow corner  in this iteration 
is  $c_2 |\Gamma^{in}_t(x,\dg)|$,  where $c_2\in O(1)$ is
the constant cost of  performing a single operation in this process. 
Observe that, if a new shadow corner of $(t,x)$ is found in this iteration,  
 it will not be considered in any subsequent iteration (Lines 28-29).
Hence, the 
 cost $c_2 |\Gamma^{in}_t(x,\dg)|$  will be incurred at most once
 for each shadow corner of $(t,x)$; that is, at most $n$ times.
 Summarizing, for each $x\in V, t\in \Z_p$ this process costs
at most $2 c_2 n |\Gamma^{in}_t(x,\dg)|$.
Hence the total cost of this process over all iterations is
$$
\sum_{x\in V, t\in \Z_p}
2\ c_2\ n\  |\Gamma^{in}_t(x,\dg)|
= 
4\ c_2\ n\ \sum_{t=0}^{p-1}
 m_t 
=
O(n m).$$

Consider now the last step of the algorithm,  of determining
if the constructed $\ag$   contains an \roo\ \uni. 
To determine all the \uni s (if any) in $\agm$ can be done by checking the degree
of each temporal node in  $\agm$, i.e., in $O(n p)$ time. To determine if
at least one of them is \roo\ can be done by a DFS traversal of  $\agm$ starting from each root 
node $(0,x)$, for a total of at most $O(n^2 + n m)$ operations.

It follows that  the total cost of the algorithm is $O(p\ n^2 + n m)$ as claimed.
\end{proof}

%-------------------
\subsection{Extensions and Improvements}
%-------------------

%-------------------
\subsubsection{Determining a Copwin Strategy}

The algorithm, as described, determines whether or not the 
arena $\dg$ (and, thus, the corresponding temporal graph $\tg$)
is copwin.  Simple additions to the algorithm
would allow it to easily determine a copwin strategy $\sigma_c$
if $\dg$ is copwin.

For any shadow edge $e=((t,x),(t+1,y)$  
let $\rho(t,x,y)$ be defined as follows:\\

1) if $e=((t,x),(t+1,y)\in E(\dg)$, then $\rho(t,x,y)=y$.\\

2) if $e=((t,x),(t+1,y)\in E(\agm)\setminus E(\dg)$, when $e$
is inserted in $\eset$, either during the Initialization or the  Iteration
phase, then $\rho(t,x,y)= z$ where
$(t+1,z)$ is the shadow cover of  $(t,y)$  determined in the 
corresponding phase of the algorithm (Line 12 if Initialization, Line 28 if Iteration).\\

Recall that, if $\dg$ is copwin, $\agm$ must contain an \roo\  \uni,
say $(t,x)$, Since $(t,x)$ is a \uni,  if the cop is located on
$(t,x)$ and the robber is located on $(t,y)$, 
by moving according to $\rho$ (starting with $\rho(t,x,y)$)
the cop will eventually capture the robber.
Since $(t,x)$ is \roo, it is reachable
from some node in $G_0$, say $(0,v)$; that is,  there is a 
journey $\pi((0,v),(t,x))$  from $(0,v)$ to $(T,x)$, where $[T]_p =t$.

Consider now the following strategy $\sigma_c$
for the cop: 
(1) choose as initial location $(0,v)$; (2) follow 
$\pi((0,v),(t,x))$; (3) follow $\rho$.
Using this strategy, the cop will eventually 
 capture the robber.

%-------------------
\subsubsection{Improvements}

The time costs of the algorithm can be reduced by
 simple modifications  and/or by exploiting 
properties of the temporal graphs.

First of all observe that the algorithm can be made to stop as soon as
a temporal node becomes a  \uni\  (e.g., testing if  $(t,x)$ is 
a \uni\ in Line 22) 
possibly  reducing 
the overall cost with the early termination.

Observe next that some of the costs of the algorithm can be reduced and 
some of its processes simplified if $\tg$ has special properties.
Consider for  example the properties of {\em reflexivity} and
{\em temporal connectivity} with respect to the last step of the algorithm,   determining
 if  $\agm$   
contains an \roo\ \uni.

\begin{property} 
\label{special} \ \\
(i) If $\tg$ is sourceless  then  every temporal node of $\dg$ is \roo.\\
\noindent (ii) Let $\tg$ be reflexive and temporally connected. If an augmented arena
$\ag$ of $\tg$ contains a \uni,   every temporal node of $\agm$ is an \roo\ \uni.
\end{property}

\begin{proof}
  (i)  Let $(t,u)$ be a temporal node of $\dg$; if $t = 0$, then $(t,u)$ is  \roo\ by definition. Let  $t > 0$; since, by assumption, no temporal node of $\tg$ is a  source, there exists a sequence of edges  
 $e_j = ((t-j-1, u_{j-1}), (t-j, u_j))\in E(\dg)$ where  $0\leq j\leq t$ 
 and $u_0 = u$. In other words there is a journey from $(0, u_{0})$ to $(t,u)$, i.e., 
 $(t,u)$ is \roo. \\
 (ii) Let $\tg$ be reflexive and temporally connected and contain a \uni, say $(i,v)$.
Since it is reflexive, then  every snapshot is sourceless, thus, by part (i) of this property,    every temporal node of $\dg$ is \roo. 
Furthermore, it is possible for the cop to wait at any vertex for any amount of time. 
This implies that, if the cop reaches vertex $v$ at any time $t$, it can just wait there until time $[t']_p=i$; in other words, all the temporal nodes $(j,v)$, $ 0\leq j < p$, are \uni.  
It also implies that,
 if the cop can reach  a \uni\  from a temporal node   $(t,u)$,
 also that node is \uni, and so are all its other temporal instances   $(j,u)$, $ 0\leq j < p$. Since $\tg$ is temporally connected, every vertex $u\in V$ is reachable at some time  starting at any time from every other vertex. Thus, the property holds.
\end{proof}

The standard game in a static graph assumes the graph to be reflexive and connected. Hence, if its extension to a periodic graph likewise assumes the graph to be reflexive and temporally connected, by Property~\ref{special}, the last step of the algorithm would consists of just testing if the degree of an arbitrary node in $\agm$ is $n$. That is, instead of $\bigo{nm}$ operations, a single one suffices.

%-------------------
\subsubsection{Game Variations}

All our results are established for the unified version of the game.
Therefore. all the caracterization properties and algorithmic results
  hold  for  
 the standard  and for the directed games studied in the literature,
  both when the players are restless and when they are not.
  
  They 
hold also for all those settings (and, thus, variants of the game)
not considered in the literature,
 where there is a  mix of vertices:  those where the players must leave and
those where  the players can wait; furthermore such a mix
might be time-varying (i.e., different in every round).

%-------------------
\section{$k>1$ Cops \& Robber}
\label{sec:k>1}

In this section, we consider the {\em C\&R} game in periodic temporal graphs
when there are  $k > 1$ cops. 
As a first step, we introduce some terminology, and we then show how to
extend the definitions and most of the previous results to this more complex setting.

%------------------------
\subsection{Terminology}
%------------------------

%----------------------
\subsubsection{Configurations}

Given a finite {\em multiset} (mset) $X$, we shall denote by $||X||$ its {\em wordcount}
(i.e., the number of its elements), by  $X^*$ its
{\em supporting} set (i.e., the set of its distinct elements), and by $\kappa(x)$ the
{\em count}  (i.e., the multiplicity) of $x\in X^*$ in $X$.

Given a finite set $P$ of size $|P|=n$,  a positive integer $k$, and
a mset $X$, we shall denote by 
$X\sqsubseteq_k P$ the facts that $X^* \subseteq P$ and
 $||X||=k$;  we shall say that $X$ is a mset over $P$
 with wordcount $k$, and denote by $[P]^k$ the collection of
 all such multisets. 

Given an arena  $\dg = (\Z_p\times V, E(\dg))$ and a  multiset  $X\sqsubseteq_k V$,
we shall denote
by $\Gamma_{[t]_p}(X,\dg) =  \bigcup_{1\leq j \leq k}\ \Gamma_{[t]_p}(x_j,\dg)$
the set of all the outneighbours  of the vertices $x_j\in X$ in $G_{[t]_p}$.

A \emph{configuration} is a triple $\conf(t,{C}, r)\in \Z^+\times [V]^k\times V$,
where the multiset $C= \langle c_1,...,c_k\rangle \sqsubseteq_k V$ denotes the positions of the cops and $r\in V$ the position of the robber
at the beginning of  round $t\in\Z^+$. In particular, $C$ specifies the number $\kappa(c)$ of cops in vertex $c\in C^*$ at the beginning of that round.

The {\em configuration graph} $\confgr = (V(\confgr),E(\confgr))$ of $\dg$, defined by
\begin{align*}
V({\confgr}) &=\{(t,{C},r)  |\  t\in\Z^+; ([t]_p,r)\in V(\dg);  \forall c_i\in{C},  
 ([t]_p,c_i) \in V(\dg)\} \\
E({\confgr}) &=\{((t,{C=\langle c_1,...,c_k\rangle},r),(t+1,{C'=\langle c'_1,...,c'_k\rangle},r')) |\  t\in\Z^+; 
c'_i\in\npg{[t]_p}{c_i}{\dg}; r\notin {C} \}
\end{align*}
is still acyclic;  the {\em source} nodes  are those with $t=0$,
 the {\em sink} nodes are those with $r\in {C}$.

A playing {\em strategy for the cops} is
any function $\cs: V({\confgr}) \rightarrow  [V]^k$
where, for every $(t,{C},r)\in V({\confgr})$,
if $r\notin{C}$ then  $((t,{C},r), (t+1,\cs(t,{C},r))\in E({\confgr})$, else $\cs(t,{C},r)={C}$.
It  specifies where the cops should move in round $t$ 
if they are at $([t]_p,C)$, the robber is at $([t]_p,r)$, 
and it is the cops'  turn to move.
A  playing  strategy $\rs$ for the robber is defined in a similar way.

A configuration $\conf(t,{C},r)$ is said to be {\em $k$-copwin}
if there exists a strategy $\cs$  such that, 
starting from  $\conf(t,{C},r)$,
 the cops  win the game  regardless of the strategy $\rs$ 
  of the  robber; such a strategy $\cs$ is said to be 
 {\em $k$-copwin for}  $\conf(t,{C},r)$.
 A strategy $\cs$ is said to be {\em $k$-copwin}  if 
there exists a ${C}$ such that
$\cs$ is $k$-copwin  for  $\conf(0,{C},r)$ for all $r\in V$.
 
If a $k$-copwin strategy exists, then $\tg$ and
 its arena $\dg$ are said to be {\em $k$-copwin}; else
 they are {\em $k$-robberwin}.

%------------------------
\subsubsection{Directed Multi-Hypergraph}

Given a set $V$ of vertices, a  {\em directed multi-hypergraph} 
(dmh) on $V$ 
is a pair  $H=(V, {\cal E}(H))$
where ${\cal E}(H)$ is  a set of ordered
pairs, called
 {\em hyperedges}, of non-empty multisets over $V$.
 For  hyperedge $e = (V_e^-,V_e^+)\in {\cal E}(H)$,  $V_e^-$ and $V_e^+$ are called the
{\em in-set} (or {tail})  and the {\em out-set} (or {head}) of   $e$, respectively;
 $|| V_e^- ||$ and $|| V_e^+ ||$ are  called the in-size 
and the out-size of $e$, respectively.

A directed multi-hypergraph $H$ is {\em homogeneous} if all its hyperedges have the same
in-size, and the same out-size, i.e., $\forall e,e'\in {\cal E}(H),  || V_e^-|| = || V_{e'}^-||$
and $|| V_e^+|| =|| V_{e'}^+||$. 
Since the focus of our study is on games with $k$ cops and a single
robber, in the following   we shall  consider only homogeneous 
directed multi-hypergraphs 
 where, for  every  hyperedge $e\in{\cal E}(H)$,
$|| V_e^-|| =k $ and $||V_e^+|| =1$. For simplicity of notation, since $V_e^+$ is a singleton, we shall denote $V_e^+$ directly by its element. Moreover, we shall denote $[V]^1$ simply by $V$.

%------------------------
\subsubsection{Augmented $k$-Arenas}

We can extend the notions of   arena  and augmented arenas from   graphs to 
hypergraphs
in a  direct way.

To each arena $\dg = (V, E(\dg))$ and integer $k\geq 1$, there corresponds a unique
 hypergraph  $\dg^k=(V,{\cal E}(\dg^k))$, called  $k${\em -arena}   defined 
as follows for all  $t\in \Z_p$ and all $x,y\in V$:
$$((t,x),([t+1]_p,y))\in E(\dg) \iff 
\forall X \in [V]^k : x\in X, ((t,X),([t+1]_p,\langle y\rangle)) \in 
{\cal E}(\dg^k) .$$

\noindent Observe that, like the arena $\dg$,  the $k$-arena $\dg^k$ is composed of $p$ 
slices $S_t(\dg^k)$, $t\in\Z_p$; and that  $\dg^1$ is precisely  $\dg$.

\begin{definition}
An {\em augmented $k$-arena} of $\dg = (V, E(\dg))$
is any   hypergraph 
$\ag^k=(V, {\cal E}(\ag))$ where: \\
(1) ${\cal E}(\dg^k)\subseteq  {\cal E}(\ag^k)$;\\
(2) for  every $t\in \Z_p$ and hyperedge $e \in  {\cal E}(\ag^k)$,  
 the configuration 
  $\conf(t,V_e^-,V_e^+)$ is $k$-copwin in $\dg$.
\end{definition}

\noindent We shall refer to the hyperedges of the augmented $k$-arena
$\ag^k$ of $\dg$ as {\em shadow hyperedges}. 

Let  $\mathbb{A}^k(\dg)$ denote the set of augmented $k$-arenas  of $\dg$.
Observe that, by definition, $\dg^k\in \mathbb{A}^k(\dg)$.
Further observe the following.

\begin{property}
\label{prop-k-closed}
The partial order $(\mathbb{A}(\dg^k),\subset)$ induced by hyperedge-set inclusion 
on $\mathbb{A}(\dg^k)$ is a complete lattice. Hence  $(\mathbb{A}(\dg^k),\subset)$ has a maximum which we denote by $\ag^{k}_{max}$.
\end{property}
\begin{proof}
It follows from the fact that, by definition
of augmented $k$-arena, the set  $\mathbb{A}(\dg^k)$  
is closed under the union of augmented $k$-arenas.
\end{proof}

\begin{definition}
Let $\ag^k$ be an augmented $k$-arena of $\dg$. 
A temporal node $(t,y)$ is a \emph{shadow $k$-corner} of a multiset $X= \langle (t+1,x_1),..., (t+1,x_k)\rangle \sqsubseteq_k V(\dg)$ with $([t+1]_p,y)\notin X$, if
\[
\npg{t}{y}{\dg} \subseteq \Gamma_{[t+1]_p}({X},{\ag^k}).
\]
The multiset  $X$ will then be called the \emph{shadow $k$-cover} of $(t,y)$.
\end{definition}

%------------------------
\subsection{$k$-Properties}
%------------------------

The equivalent of the three basic properties of augmented arenas
established in Section~\ref{sect:CopwinPeriodicGraphs}, namely Theorems~\ref{thm:univ_temp_node}, \ref{thm:ccg_add_edge} and~\ref{thm:maximal}, can be shown to
hold also when $k>1$, following
analogous proof arguments.

For simplicity, we shall call a multiset
$X\sqsubseteq_k V(S_t)$ a {\em $k$-temporal node}.
A $k$-temporal node $X = \langle (t,x_1),(t,x_2),...,(t,x_k)\rangle\sqsubseteq_k V(S_t)$ is said to be a  \emph{$k$-\uni} if $\Gamma_t(X,\dg) =  V$.
It is said to be {\em $k$-\roo}\ if 
there exists a multiset 
$U= \langle (0,u_1),(0,u_2),...,(0,u_k)\rangle \sqsubseteq_k  V(\dg)$
such that there is a journey $\pi_j$ from $(0,u_j)$ to $(t,x_j)$ for all $1\leq j\leq k$.

The following theorem is a generalization of Theorem~\ref{thm:univ_temp_node} for $k$ cops.
\begin{theorem}
\label{thm:k_universal}
{\em ($k$-Characterization Property)}\\
An arena $\dg$ is $k$-copwin if and only  if 
$\ag^{k}_{max}$ contains a 
$k$-\roo\  $k$-\uni.
\end{theorem}
\begin{proof}
({\bf  if})
Let $\ag^{k}_{max}$ contain a $k$-\roo\  $k$-\uni\ $U=\langle (t,u_1),(t,u_2),...,(t,u_k)\rangle\sqsubseteq_k V(\dg)$. By definition of $k$-\uni, $\Gamma_t(U,\dg^k) =  V$.
Thus, by definition of $k$-augmented arena, 
for every $v\in V$ the configuration $\conf(t,U,v)$ is $k$-copwin, i.e.,
there is a $k$-copwin strategy $\cs$ from $\conf(t,U,v)$. 

Since $U$ is $k$-\roo, there exists a $k$-temporal node  
$X= \langle (0,x_1),(0,x_2),..., (0,x_k)\rangle \sqsubseteq_k  V(\dg)$ such that there is a journey $\pi_j$ from $(0,x_j)$ to $(t,u_j)$ for all $1\leq j \leq k$.
Consider now the cop strategy $\cs'$ of: (1)  initially positioning the $k$ cops $c_1,c_2,...,c_k$ on the nodes in $X$, (2) then each $c_j$ moving according to the journey $\pi_j((0,x_j),(t,u_j))$ and, once on $(t,u_j)$, (3) following the copwin strategy $\cs$ from $\conf(t,U,w)$, where $w$ is the position of the robber at the beginning of round $t$. This strategy $\cs'$ is winning for $\conf(0,X,v)$ for all $v\in V$; hence $\dg$ is $k$-copwin.

({\bf only if}) 
Let $\dg$ be $k$-copwin. We show that there must exist an augmented $k$-arena $\ag^{k}$ of $\dg$ that contains a $k$-\roo\ $k$-\uni. Since $\dg$ is copwin, by definition, there must exist some starting position $X= \langle (0,x_1),(0,x_2),..., (0,x_k)\rangle \sqsubseteq_k  V(\dg)$ for the cops such that, for all positions initially chosen by the robber, the
cops eventually capture the robber. In other words, all the configurations
$\conf(0,X,v)$ with $v\in V$ are $k$-copwin; thus the $k$-arena $\ag^{k}$ obtained 
by adding to $E(\dg)$ the set of hyperedges $\{(0,\langle x_1, x_2,..., x_k \rangle,v) | v\in V\}$
is an augmented $k$-arena of $\dg$ and the multiset $X$ is a $k$-\roo\ $k$-\uni. By Property~\ref{prop-k-closed},  $E(\ag^k)\subseteq E(\ag^{k}_{max})$ and the theorem follows.
\end{proof}

The following theorem is a generalization of Theorem~\ref{thm:ccg_add_edge} for $k$ cops.
\begin{theorem}
\label{thm:k_aug_prop}{\em ($k$-Augmentation Property)}\\
Let $\ag^k\in \mathbb{A}^k(\dg)$,
$X= \langle (t,x_1),..., (t,x_k)\rangle \sqsubseteq_k V(\dg)$,
$(t,y)\in V(\dg)$
and 
$Z= \langle (t+1,z_1),..., (t+1,z_k)\rangle \sqsubseteq_k V(\dg)$.
Assume there is a bijection $f:X\rightarrow Z$ such that
$f(t,x_i)$ is an out-neighbour of $(t,x_i)$ for all $1\leq i\leq k$.
If $(t,y)$ is a shadow $k$-corner of $Z$, 
then the arena ${\ag'}^k = \ag^k \cup (X,(t+1,y))$ is an augmented $k$-arena of $\dg$.
\end{theorem}

\begin{proof}
Let $\ag^k$ be an  augmented $k$-arena of $\dg$ and let $X= \langle (t,x_1),..., (t,x_k)\rangle \sqsubseteq_k V(\dg)$,
$(t,y)\in V(\dg)$,
$Z= \langle (t+1,z_1),..., (t+1,z_k)\rangle \sqsubseteq_k V(\dg)$.
Assume there is a bijection $f:X\rightarrow Z$ such that
$f(t,x_i)$ is an out-neighbour of $(t,x_i)$ for all $1\leq i\leq k$,
and $(t,y)$ is a shadow $k$-corner of $Z$.
The theorem  follows if $(X,(t+1,y))$ is already a hyperedge of $E(\ag^k)$.
Consider  the case where $(X,(t+1,y))\notin E(\ag^k)$.
Since $(t,y)$ is a  shadow corner of $Z$, then for every $w\in \npg{t}{y}{\dg}$ we have that $(Z,(t+2,w))\in E(\ag^k)$, i.e.,  
$\conf(t+1,Z,w)$ is winning for the cops.
If the cops move from $X$ to $Z$ using $f$
when the robber is on $(t,y)$, then regardless of the robber's move, 
the resulting configuration would be winning for the cops.
In other words, $\conf(t,X,y)$ is a winning configuration for the cops.
It follows that ${\ag'}^k = \ag^k \cup \set{(X,(t+1,y))}$ is an  augmented arena  of $\dg^k$.
\end{proof}

To be able to 
determine whether the current augmented $k$-arena  of $\dg$ is  indeed $\ag^{k}_{max}$, we can use the following theorem.
This is a generalization of Theorem~\ref{thm:maximal} for $k$ cops.
\begin{theorem}\label{thm:k-maximal}
{\em ($k$-Maximality Property)}\\
Let  $\ag^k\in \mathbb{A}^k(\dg)$.  Then   $\ag^k=\ag^{k}_{max}$  if and only if, for every hyperedge $(X,(t+1,y))\notin E(\ag^k)$, where $X\sqsubseteq_k V(S_t)$, there exists no $Z \sqsubseteq_k V(S_{t+1})$ for which (1) there exists a bijection  $f:X\rightarrow Z$ where $f(t,x_j)$ is an out-neighbour of $(t,x_j)$ for all $1\leq j\leq k$, and (2) $(t,y)$ is a shadow $k$-corner of $Z$.
\end{theorem}

\begin{proof}
{\bf (only if)}\ 
By contradiction, let  $\ag^k=\ag^{k}_{max}$ but there exists a hyperedge $(X,(t+1,y)\notin E(\ag^k)$ where $X\sqsubseteq_k V(S_t)$; 
and a $k$-temporal node $Z \sqsubseteq_k V(S_i)$ for which (1) there exists bijection  $f:X\rightarrow Z$ where $f(t,x_j)$ is an out-neighbour of $(t,x_j)$ for all $1\leq j\leq k$, and (2) $(t,y)$ is a shadow $k$-corner of $Z$. By Theorem~\ref{thm:k_aug_prop}, $\ag'^k = \ag^k \cup (X,(i+1,y))$ is an  augmented arena of $\dg$; however, $E(\ag'^k)$ contains one more edge than $E(\ag^k)$, contradicting the assumption that $\ag^k$ is maximum.

{\bf (if)}\ 
Let  $\ag^k \neq \ag^{k}_{max}$; that is, there exists $(X,(t+1,y))\in E(\ag^{k}_{max})\setminus E(\ag^k)$. By definition, the configuration $\conf(t,X,y)$ is $k$-copwin; let $\cs$ be a copwin strategy for the
configuration $\conf(t,X,y)$, i.e., starting from  $\conf(t,X,y)$, the cops win the game regardless of the strategy $\rs$ of the robber.

Let $\gamegr = (V(\gamegr),E(\gamegr))\subseteq \confgr$ be the directed acyclic graph of configurations induced by $\cs$ starting from $\conf(t,X,y)$, and defined as follows:
(1) $\conf(t,X,y)\in V(\gamegr)$; 
(2) if $\conf(t',Y,v)\in V(\gamegr)$ with $t'\geq t$ and $v\notin Y$, then, for all $w\in  \Gamma_{t'}(v,\dg)$, $\conf(t'+1,\cs(t'+1,Y,v),w) \in V(\gamegr)$ and $(\conf(t',Y,v), \conf(t'+1,\cs(t'+1,Y,v),w)))\in E(\gamegr)$.
    
Observe that in $\gamegr$ there is only one source (or root) node, $\conf(t,X,y)$, and every $\conf(t',Z,w)\in V(\gamegr)$ with $w\in Z$ is a sink (or terminal) node. Since $\cs$ is a winning strategy for the root, every node in $\gamegr$ is a copwin configuration, and every path from the root terminates in a sink node. 

Partition $V(\gamegr)$ into two sets, $U$ and $W$ where $U=\{\conf(i,Y,v) | (Y,(i+1,v))\in E(\ag^k)\}$ and $W= V(\gamegr) \setminus U$. Observe that every sink of $V(\gamegr)$ belongs to $U$; on the other hand, since  $(X,(t+1,y))\notin E(\ag^k)$ by assumption, the root belongs to $W$.
 
Given a node $\kappa=\conf(i,Y,v)\in V(\gamegr)$, let $\gamegr[\kappa]$ denote the subgraph of $\gamegr$ rooted in $\kappa$. \\

\noindent {\bf Claim.} {\em There exists  $\kappa\in  V(\gamegr)$ such that all nodes of $\gamegr[\kappa]$, except $\kappa$, belong to $U$. }\\
{\bf Proof of Claim.}
Let $P_0$ be the set of sinks of $\gamegr$. Starting from $s=0$, consider the set $P_{s+1}$ of all in-neighbours of any node of $P_{s}$; if $P_{s+1}$ does not contain an element of $W$, then increase $s$ and repeat the process. Since $(t,X,y)\in W$, this process terminates for some $s\geq 0$, and the Claim holds for every $\kappa\in P_{s+1}\cap W$. $\Box$\\

Let $(t',X',y')$ be a node of $V(\gamegr)$ satisfying the above Claim. Thus $(X',(t'+1,y'))\notin E(\ag^k)$ but, since $(i',X',y')$ is copwin, $(X',(t'+1,y'))\in \ag^{k}_{max}$. By the Claim, all other nodes of $\gamegr[(t',X',y')]$ belong to $U$, in particular the set of nodes $\{(t'+1,B,z) | B = \cs(t',X',y'), z\in\Gamma_{t'}(y',\dg)\}$. This means that, for every $z\in\Gamma_{t'}(y',\dg)$, $(t'+1,B,z)\in E(\ag^k)$. In other words, $\npg{t'}{y'}{\dg}\subseteq \npg{t'+1}{B}{\ag^k}$; that is,  $(t',y')$ is a shadow corner of $B$.

Summarizing: by assumption $\ag^k \neq \ag^{k}_{max}$; as shown, $(X',(t'+1,y'))\in E(\ag^{k}_{max})\setminus E(\ag^k)$, and $B \subseteq \npg{t'}{X'}{\dg}$ is a shadow cover of $(t',y')$; that is, $\npg{t'}{y'}{\dg}\subseteq \npg{t'+1}{B}{\ag^k}$, concluding the proof of the {\bf if} part of the theorem.
\end{proof}

 \begin{figure}
\begin{center}
\fbox{
\begin{minipage}{5.5cm}
{\sc  General Strategy ($k\geq 1$)}
  \begin{tabbing} 
1. W\=hi\=le there  is a still unexamined shadow hyperedge $e=(X,(t+1,y))\in E(\ag^k)$ do:\\
2. \>  If there are still unexamined shadow $k$-corners   covered by $X$ then:\\
3. \> \>  F\=or each such shadow $k$-corner  $(t-1,z)$  do:\\
4. \> \>  \> If  th\=ere are new  shadow hyperedges due to  $(t-1,z)$ then:\\
 5. \> \>  \>   \>            Add  them to  $\ag^k$ to be examined. \\
6. \> \>  \> Remove $(t-1,z)$ from consideration as a  shadow corners of $X$ (i.e., mark it as examined).\\
 7. \> Remove $e$ from consideration (i.e., mark it as examined).\\
 8. If there is an  \ \uni,  then $\dg$ is copwin else it is robberwin.
  \end{tabbing}
\end{minipage}
}
\end{center}
\caption{Outline of general strategy; it terminates when $\ag^k=\ag^{k}_{max}$.} \label{fig:kstrategy}
\end{figure}

%------------------------
\subsection{$k$-Copwin Determination}
%------------------------

Based on the characterization and  properties  of $k$-copwin periodic graphs established in the previous section, a  general strategy to determine if a peridic graph is $k$-copwin follows directly (see Figure~\ref{fig:kstrategy}) along the same lines of the one discussed in Section~\ref{analysis}.
The immediate  implementation of the strategy does lead to a solution algorithm with proof of correctness  and  complexity analysis following  exactly the same lines of that for $k=1$; its time complexity however does not improve  the  $O(k \ p\ n^{k+2})$  bound reported in \cite{EMSW24}.

%------------------------
\section{Conclusions}
%------------------------

 In this paper we have provided a  complete characterization
 of  copwin {\em periodic} temporal graphs, 
 establishing several basic properties on the 
 nature of a copwin  game in such graphs.

These characterization results are {\em general}, 
in the sense that they  
do not rely on any  assumption on properties
such as connectivity, symmetry, reflexivity
held (or not held) by the individual static graphs 
in the periodic sequence defining the temporal graph.

These results  have been established for the {\em unified} version of  the game,
which includes the standard  undirected and directed  versions
of the game, both in the original and restless variants, as well as new variants never studied before.

Based on these results,  we have also designed an
algorithm that determines if a  temporal graph with period $p$
is copwin in time    $O(p\ n^2 + n m)$,
 where $m$ is the number of edges in the arena,
 improving the (then) existing bounds for periodic and static graphs.
  Observe that it has been recently shown in \cite{EMSW24} that our bound can be reached  also by using the reduction to reachability games of  \cite{ErS20}.

In the case  $k>1$ of multiple cops,  by shifting from a representation in terms of directed graphs to one in terms of directed multi-hypergraphs, we proved that all the fundamental properties of augmented arenas
established for $k=1$ continue to hold,
providing a complete characterization of $k$-copwin periodic graphs.

The established results for $k>1$ lead directly to a solution strategy to determine if a periodic temporal graph is $k$-copwin; however, the immediate straightforward implementation of the strategy  appears to achieve a less efficient results that the reported  $O(k \ p\ n^{k+2})$  bound reported in \cite{EMSW24}. 
An outstanding open problem is whether, using the characterization properties established here, it is possible to match if not improve the existing bound.

Other than the algorithmic aspects, a major open research direction is 
the study of structural properties of copwin temporal graphs, extending to the temporal realm the extensive investigations  carried out on static graphs. This investigation has just started  \cite{DeFSS23b}.

\ \\

\noindent {\bf Acknowledgments}\\
\noindent This research has been supported in part by the Natural Sciencies and Engineering Research Council of Canada (NSERC) under the Discovery Grant program.
Some of these results have been presented at the 30th International Colloquium
on Structural Information and Communication Complexity \cite{DeFSS23}.

\bibliographystyle{plain}
\bibliography{SiroccoJ}

\end{document}